\pdfoutput=1
\RequirePackage{lineno}
\newif\ifpreprint%
\preprintfalse%
\ifpreprint%
	\documentclass[preprint,english,aps,prb,floatfix,amssymb,superscriptaddress,a4paper]{revtex4}
\else%
	\documentclass[twocolumn,english,aps,prb,floatfix,amssymb,superscriptaddress,a4paper]{revtex4}
\fi%
\usepackage{amsmath}
\usepackage{siunitx}
\usepackage{dsfont}
\usepackage{color}
\usepackage{amssymb}
\usepackage{graphicx}
\usepackage{braket}
\usepackage{units}
\usepackage{chemformula}
\usepackage{amssymb}
\usepackage{mathtools}
\usepackage{physics}
\usepackage{braket}
\usepackage{bibunits}

\usepackage{hyperref}
\hypersetup{
  colorlinks = true,
  hypertexnames=true
}
\newcommand{\ssm}{\scriptscriptstyle\rm}

\renewcommand{\theta}{\vartheta}
\renewcommand{\phi}{\varphi}

\newcommand{\RN}[1]{%
  \textup{\uppercase\expandafter{\romannumeral#1}}%
}

\begin{document}
\ifpreprint%
	\linenumbers%
\fi%

\title{Fragile topology and flat-band superconductivity in the strong-coupling regime}

\author{
Valerio Peri}
\email{periv@phys.ethz.ch}
\affiliation{
Institute for Theoretical Physics, ETH Zurich, 8093 Z\"urich, Switzerland
}

\author{
Zhida Song}
\affiliation{
Department of Physics, Princeton University, Princeton, New Jersey 08544, USA
}

\author{
B. Andrei Bernevig}
\affiliation{
Department of Physics, Princeton University, Princeton, New Jersey 08544, USA
}

\author{
Sebastian D. Huber}
\affiliation{
Institute for Theoretical Physics, ETH Zurich, 8093 Z\"urich, Switzerland
}

\let\oldaddcontentsline\addcontentsline
\renewcommand{\addcontentsline}[3]{}
\begin{bibunit}[apsrev4-2]
\begin{abstract}
In flat bands, superconductivity can lead to surprising transport effects. The superfluid ``mobility'', in the form of the superfluid weight $D_s$, does not draw from the curvature of the band but has a purely band-geometric origin. In a mean-field description, a non-zero Chern number or fragile topology sets a lower bound for $D_s$, which, via the Berezinskii-Kosterlitz-Thouless mechanism, might explain the relatively high superconducting transition temperature measured in magic-angle twisted bilayer graphene (MATBG). For fragile topology, relevant for the bilayer system, the fate of this bound for finite temperature and beyond the mean-field approximation remained, however, unclear. Here, we use numerically exact Monte Carlo simulations to study an attractive Hubbard model in flat bands with topological properties akin to those of MATBG. We find a superconducting phase transition with a critical temperature that scales linearly with the interaction strength. We then investigate the robustness of the superconducting state to the addition of trivial bands that may or may not trivialize the fragile topology. Our results substantiate the validity of the topological bound beyond the mean-field regime and further stress the importance of fragile topology for flat-band superconductivity.

\end{abstract}

\date{\today}

\maketitle

Whenever the single particle's kinetic energy does not depend on momentum, non-interacting electrons, if not in a topological state, are strictly localized. Nevertheless, these seemingly inert systems exhibit intriguing transport phenomena in the presence of many-body effects. A paradigmatic example is the onset of unconventional superconductivity, where mobile coherent electron pairs emerge from an insulating high-temperature state \cite{Khodel:1990,Kopnin:2011,Volovik:2013,Iglovikov:2014,Tovmasyan:2016,Kauppila:2016,Kobayashi:2016,Lothman:2017,Ojajarvi:2018,Kumar:2019,Nunes:2020,Swain:2020a} under the influence of strong electron-electron interactions. The interest in this flat-band superconductivity surged after its observation in magic-angle twisted bilayer graphene (MATBG) \cite{Cao:2018}. Since then, signatures of zero-resistance states have been reported in other flat-band van der Waals systems such as twisted double-bilayer graphene \cite{Shen:2020,Cao:2020,Liu:2019a}, twisted trilayer graphene \cite{Tsai:2019}, ABC--trilayer graphene \cite{Chen:2019}, and bilayer \ch{WSe_2} \cite{Wang:2019d}. 

Superconductivity arises from the interplay of two different energy scales: the effective electron-electron attractive interaction $\lvert U \rvert$ and the bandwidth $W$. A vanishing bandwidth maximizes the density of states $ n_0(\epsilon_F)$ at the Fermi energy and the Bardeen-Cooper-Schrieffer (BCS) theory predicts $T_{c,\text{BCS}} \propto \lvert U \rvert n_0(\epsilon_F)$ in the flat-band limit $\lvert U \rvert \gg W$ \cite{Khodel:1990,Kopnin:2011,Volovik:2013}. While the BCS theory might seem unsuitable to treat systems lacking a well-defined Fermi surface, the BCS wave function turns out to be an exact zero-temperature ground state for certain flat bands with local attractive interactions \cite{Peotta:2015,Tovmasyan:2016}. Nevertheless, the validity of the BCS theory at finite temperature is questionable and one needs to be careful in exploring the strong-coupling regime \cite{Hofmann:2019, Wang:2020}. Moreover, while the BCS theory captures the formation of electronic pairs, their phase fluctuations are know to be crucial in two-dimensional (2D) superconductors \cite{Emery:1995a}.

Phase coherence emerges via the Berezinskii-Kosterlitz-Thouless (BKT) mechanism \cite{Kosterlitz:1973,Moreo:1991,Assaad:1994}. Within the BKT theory, the fraction of electrons condensed into coherent bound pairs is captured by the superfluid weight $D_s(T)$. The universal jump in this quantity determines the transition temperature $T_{c}$: $T_{c}=\pi D_s^-/2$, where $D_s^-$ is the superfluid weight at the critical temperature approached from below \cite{Nelson:1977}. 

The Ginzburg-Landau theory for conventional superconductors predicts $D_s(T=0)\approx e^2n_s/m^*$, where $n_s$ is the amplitude of the superconducting order parameter, and $m^*$ is the effective band mass \cite{Scalapino:1992}. Exactly flat bands have an infinite effective mass, $m^*=\infty$. Hence, a vanishing bandwidth seems detrimental to phase coherence. Therefore, one would expect phase fluctuations to completely disrupt superconductivity in dispersionless bands. However, this conclusion neglects other band properties that are not captured by a simple effective mass approximation.

The presence of a further contribution to the superfluid weight is now well-established in the mean-field approximation \cite{Peotta:2015,Tovmasyan:2016,Julku:2016,Liang:2017}. This additional term has a band-geometric origin and is proportional to the Fubini-Study metric of the occupied bands \cite{Peotta:2015}. Lower bounds for $D_s(T=0)$ have been formulated both for bands with a non-zero Chern number \cite{Peotta:2015} as well as for two bands characterized by fragile topology \cite{Xie:2020}. 

While the bound in terms of the Chern number and its influence on the $T\neq 0$ physics has been recently investigated numerically in the strong-coupling regime \cite{Hofmann:2019}, no such analysis has been performed for the case of fragile topology. The latter is particularly relevant since the single-particle nearly flat bands of MATBG have zero Chern number but non-trivial fragile topology \cite{Lu:2020,Lian:2018,Po:2019,Song:2019a,Ahn:2019,Zaletel:2019}. In the current manuscript, we fill this gap by studying a concrete flat-band model with fragile topology via exact numerical methods.

We first review the concept of fragile topological bands and introduce the concrete model used in this study. We then compare the superfluid weight obtained from quantum Monte Carlo simulations to the zero-temperature mean-field topological bound. To further establish the importance of fragile topology, we investigate the fate of the superconducting state under the addition of trivial bands. Finally, we analyze the properties of the normal state above the superconducting phase transition. 

Fragile Bloch bands represent a flavor of symmetry protected topological insulators (TIs) as these bands cannot be represented by translationally and lattice temporal-spatial symmetric, exponentially localized Wannier functions. However, the addition of a trivial Bloch band can resolve this obstruction, contrary to the stable TIs \cite{Po:2018,Bouhon:2018,Bradlyn:2019,Ahn:2019,Else:2019}. In the case of MATBG, the protecting symmetry  is $C_{2z}\mathcal T$, where $C_{2z}$ is a $180\si{\degree}$ rotation around the out-of-plane axis $\hat{z}$, and $\mathcal T$ is the bosonic time-reversal symmetry that acts as complex conjugation. For two occupied bands with $C_{2z}\mathcal T$ symmetry, it is possible to introduce a $\mathbb{Z}$-classification based on the Euler class, $e_2$, of real orientable bundles \cite{Song:2019a,Xie:2020,Ahn:2019,Bouhon:2018,Bouhon:2020}. In particular, Ref.~[\onlinecite{Xie:2020}] showed that, in the mean-field approximation, a non-trivial Euler class provides a lower bound on $D_s(T=0)$. The geometric contribution to the superfluid weight of MATBG has since then been further discussed in \cite{Xie:2020,Julku:2020,Hu:2019a}.

To investigate the robustness of the bound for fragile bands in the strong-coupling regime, we follow Ref.~[\onlinecite{Hofmann:2019}] and consider an attractive Hubbard model which lends itself to numerically exact auxiliary-field quantum Monte Carlo simulations \cite{Hirsch:1983,Blankenbecler:1981,Bercx:2017}. 

\begin{figure}[t]
\includegraphics[]{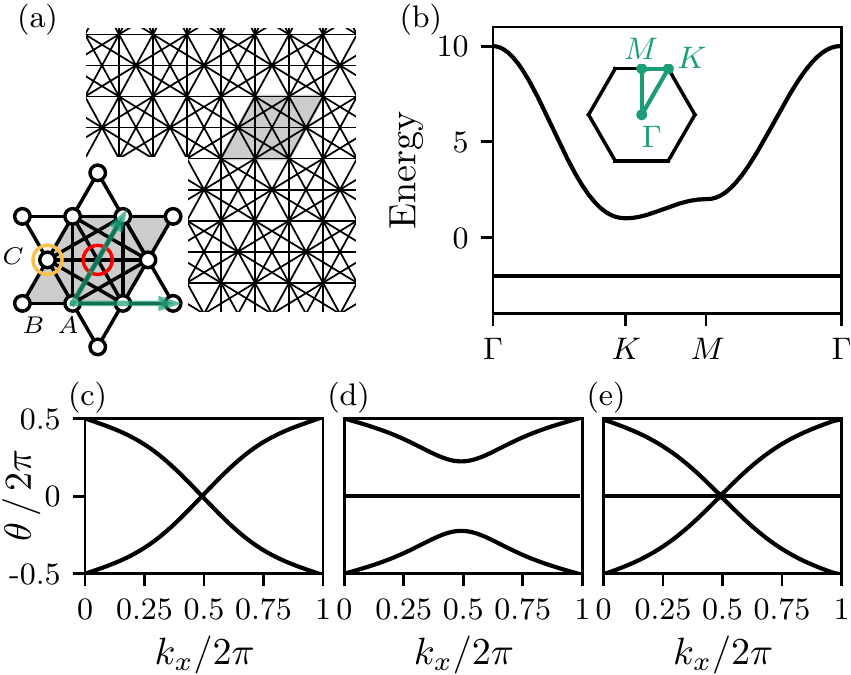}
\caption{\label{fig:fig1} (a) Kagome-3 lattice model with the unit-cell area shaded in gray. The inset shows the details of one plaquette. The red circle highlights the $1a$ Wyckoff position of the lattice, while the yellow circle the $1b$ Wyckoff position. Note that, although the full point group of the model is $p6mm$, we refer to the real space high-symmetry-points of its subgroup $p2$ \cite{supp}. The basis vectors $\mathbf{a_1}$ and $\mathbf{a_2}$ are represented by the green arrows. The three inequivalent sublattices $A$, $B$ and $C$ are also highlighted. (b) Single-particle spectrum of the model along high-symmetry lines of $p6mm$. The high-symmetry lines and the first Brillouin zone (BZ) are shown in the inset. The flat bands at $\epsilon=-2$ is doubly degenerate. (c) Wilson loop spectrum of the two flat bands of the kagome-3 model. (d) Wilson loop spectrum of the three lowest bands of the model with an additional $s$ orbital at $1a$ Wyckoff position. (e) Same as (d) but with an $s$ orbital at $1b$ Wyckoff position.}
\end{figure}

We focus on a particular 2D lattice model, known as kagome-3 \cite{Balents:2002,Bergman:2008,Rhim:2019}, which represents a minimal instance of a flat-band system characterized by fragile topology. The lattice, its basis vectors $\mathbf{a_1}=(1,0)$ and $\mathbf{a_2}=(1/2,\sqrt{3}/2)$, and the three inequivalent sublattices in the unit cell are shown in Fig.~\ref{fig:fig1}(a). We study the Hamiltonian:
\begin{align}
	\label{eq:model}
	H&=H_{\text{kin}}+H_{\text{int}},\\
	\label{eq:modelK}
	H_{\text{kin}}&=\sum_{i,j,\sigma} t_{ij} c_{i\sigma}^\dagger c^{\phantom{\dagger}}_{j\sigma}-\mu\sum_i\left(n_{i\downarrow}+n_{i\uparrow}\right),\\
	H_{\text{int}}&=-\lvert U \rvert\sum_i \left(n_{i\uparrow}-\frac{1}{2}\right)\left(n_{i\downarrow}-\frac{1}{2}\right),
	\end{align}
where $c^{\phantom{\dagger}}_{i\sigma}$ is the fermionic annihilation operator, and $n_{i\sigma}=c_{i\sigma}^\dagger c^{\phantom{\dagger}}_{i\sigma}$ counts the number of electrons on site $i$ with spin $\sigma=\{\uparrow,\downarrow\}$. $\lvert U \rvert$ is the electron-electron interaction strength, $\mu$ the chemical potential, and $t_{ij}$ the hopping parameter between site $i$ and $j$. In the remainder, we set all hopping terms to unity. 

The single-particle physics encoded in Eq.~\eqref{eq:modelK} is particularly appealing. The spectrum of the model has two degenerate flat bands at $\epsilon(\mathbf{k})=-2$ and a third dispersive band $\epsilon(\mathbf{k})=4+2\left[\cos{\mathbf{k}\cdot \mathbf{a_1}}+\cos{\mathbf{k}\cdot \mathbf{a_2}}+\cos{\mathbf{k}\cdot (\mathbf{a_1}-\mathbf{a_2})}\right]$, cf. Fig.~\ref{fig:fig1}(b). The smallest gap $\delta=3$ between the dispersive and flat bands is attained at the momentum point $K=(2\pi/3,2\pi/\sqrt{3})$ \cite{supp}. Note that the model possesses both spinful time-reversal symmetry and spin $S^z$ conservation. These features allow us to study the topological properties by computing the Wilson loop operators of each spin sector independently \cite{Peotta:2015,Xie:2020,supp}. As shown in Fig.~\ref{fig:fig1}(c), the winding in the Wilson loop spectra of the two flat bands establishes their topological nature with a non-trivial Euler class $\lvert e_2\rvert=1$ protected by $C_{2z}\mathcal T$ \cite{Ahn:2019,Bouhon:2018,supp}. The winding of the spectrum is removed by the addition of a trivial band, as shown in Fig.~\ref{fig:fig1}(d), confirming the presence of fragile topology. The topological properties of the kagome-3 flat bands are thus akin to those of the single-particle bands of MATBG \cite{Lu:2020,Lian:2018,Po:2019,Song:2019a,Ahn:2019,Zaletel:2019}. 

\begin{figure*}[t]
\includegraphics[]{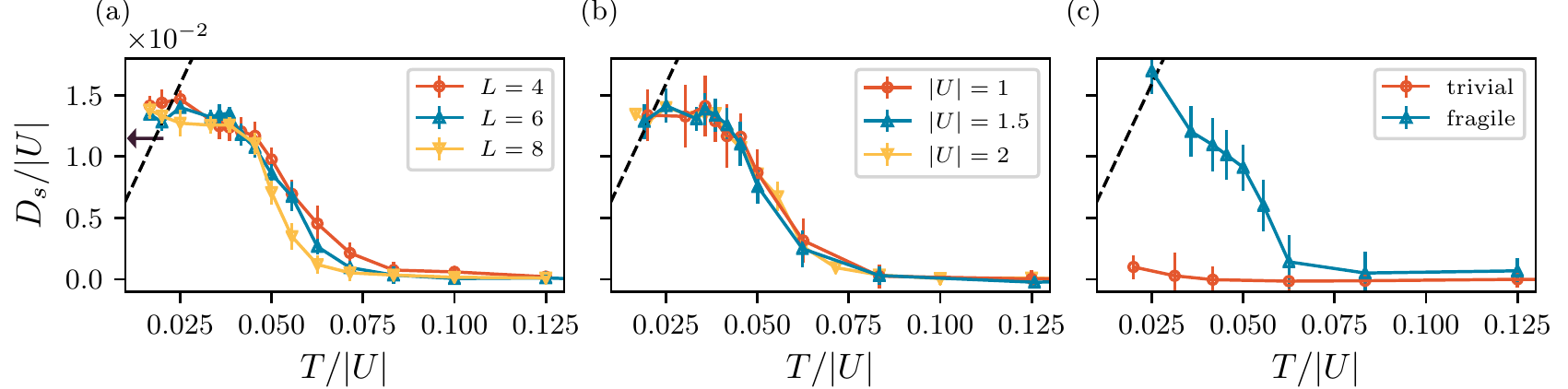}
\caption{\label{fig:fig2} Superfluid weight $D_s(T)$ for the attractive Hubbard model with interaction strength $\lvert U \rvert$. The crossing of $D_s$ with the dashed line $2T/\pi$ indicates the BKT transition, where the superconducting transition occurs. (a) Different systems sizes $L\times L$, with $L=4,6,8$ and $\lvert U \rvert =2$. The arrow on the $y$ axis represents the mean-field topological lower bound for $D_s(T=0)$ \cite{supp}. (b) Different interaction strengths $\lvert U \rvert = 1,1.5,2$ in a $6\times 6$ system. In both (a) and (b) the kagome-3 model is considered. (c) Results for the four-band models for $\lvert U \rvert=2$ and $L=6$. The trivial model has an additional $s$ orbital at Wyckoff position $1a$, cf. Fig.~\ref{fig:fig1}(a). The model with fragile topology has instead an additional $s$ orbital at Wyckoff position $1b$, cf. Fig.~\ref{fig:fig1}(a).}
\end{figure*}

To compute $D_s(T)$, we introduce an external electromagnetic field via its electromagnetic potential $\mathbf{A}$ and Peierls substitution:
	$t_{ij}\rightarrow t_{ij}\,\exp[i\mathbf{A}\cdot(\mathbf{r}_i-\mathbf{r}_j)]=t_{ij}(\mathbf{r})$, 
where $\mathbf{r}_i$ is the position of the site $i$ and $\mathbf{r}=\mathbf{r}_i-\mathbf{r}_j$. We can then expand $H(\mathbf{A})$ up to second order in $\mathbf{A}$:
\begin{equation}
H(\mathbf{A})=H+ j^p_\mu A_\mu+\frac{1}{2}T_{\mu\nu}A_\mu A_\nu,
\end{equation}
where $j^p_\mu$ is the paramagnetic current operator and $T_{\mu\nu}A_\nu$ is the diamagnetic one. These operators are defined as
\begin{equation}
	j^p_\mu=\sum_{ij,\sigma}\frac{\partial t_{ij}(\mathbf{r})}{\partial r_\mu}c^\dagger_{i\sigma}c^{\phantom{\dagger}}_{j\sigma},
\end{equation}
and 
\begin{equation}
	T_{\mu\nu}=\sum_{ij,\sigma}\frac{\partial^2 t_{ij}(\mathbf{r})}{\partial r_\mu\partial r_\nu}c^\dagger_{i\sigma}c^{\phantom{\dagger}}_{j\sigma}.
\end{equation}
The superfluid weight characterizes the zero-frequency, long-wavelength response to the external field, $j_\mu=D_{s,\mu\nu}A_\nu$. It is given by
\begin{equation}
	\label{eq:boundWeight}
	D_{s,\mu\nu}=\frac{1}{4}\left[\langle T_{\mu\nu} \rangle-\Lambda_{\mu\nu}(k_\parallel=0,k_\perp\rightarrow 0,i\omega_m=0)\right],
\end{equation}
where $k_{\parallel (\perp)}$ is the momentum component parallel (perpendicular) to $\mathbf{A}$, and $\langle\cdot\rangle$ represents the expectation value over the many-body ground state of Eq.~\eqref{eq:model} at temperature $T$. Here, $\Lambda_{\mu\nu}(\mathbf{k},\omega)$ is the current-current correlator:
\begin{equation}
\Lambda_{\mu\nu}(\mathbf{k},i\omega_m)= \int_0^\beta \text{d}\tau e^{i\omega_m \tau}\left\langle \left[j^p_\mu(\mathbf{k},\tau),j^p_\nu(-\mathbf{k},0) \right] \right\rangle,
\end{equation}
with $\omega_m=2\pi mT$, $m\in \mathbb Z$, and $\beta=1/k_{\ssm B}T$ the inverse temperature. 
In the reminder, we consider $D_s=D_{s,xx}$ and a gauge potential $\mathbf{A}=A\mathbf{\hat{x}}$.


The quantum Monte Carlo simulations grant access to the superfluid weight $D_s(T)$ in the strong-coupling regime at finite temperature \cite{supp}. We perform simulations in the grand canonical ensemble, where the chemical potential $\mu$ controls the filling $\nu$ of the system. We carefully tune $\mu(T)$ to ensure $\nu=1/3$, i.e., two electrons per unit cell and half-filling of the flat bands. We focus on a range of Hubbard interactions $\lvert U \rvert < \delta$, where $\delta$ is the energy gap between the flat bands and the dispersive one. 

First, we consider $\lvert U \rvert=2$ and lattices of different sizes $L\times L$, with $L=4,6,8$. Since each unit cell contains three inequivalent sublattices and we consider spinful electrons, the number of orbitals in these systems is $96, 216, 384$, respectively. In Fig.~\ref{fig:fig2}(a) we present the results of this analysis. The transition temperature $T_c/\lvert U\rvert \approx 0.02$ is given by the universal jump in the superfluid weight $D_s(T)$ \cite{Nelson:1977} and shows little dependence on the system's size. 

Next, we investigate the relation between $T_c$ and $\lvert U \rvert$. Since the Hubbard interaction is the only energy scale of the problem, we expect a linear relation between $T_c$ and $\lvert U \rvert$: $T_c \propto \lvert U \rvert$ \cite{Peotta:2015}. This observation is confirmed by the plot of $D_s(T/\lvert U \rvert)/\lvert U\rvert$ with $\lvert U\rvert=1,1.5,2$ for a $6\times 6$ system. The three curves lie on top of each other and confirm $T_c/\lvert U\rvert\approx 0.02$, cf. Fig.~\ref{fig:fig2}(b). These findings parallel those for flat Chern bands in the strong-coupling regime \cite{Hofmann:2019} and substantiate the onset of superconductivity in the exactly flat bands of the kagome-3 lattice.

The addition of trivial bands to a fragile topological insulator can remove the obstruction to an atomic limit and might impact the strength of the superconducting order. Therefore, we carefully assess the fate of superconductivity when further bands are considered. 

To investigate the Wannierizability under the addition of trivial bands, we resort to an analysis of the symmetry eigenvalues of inversion $C_{2z}$ \cite{supp}. This approach allows us to consider two different scenarios. First, we couple an extra $s$-orbital at the $1a$ Wyckoff position [red circle in Fig.~\ref{fig:fig1}(a)] to all adjacent sites. This additional orbital gives rise to an $A_{1a}$ band that trivializes the flat bands of the original model, see \cite{supp} for a detailed analysis.  A fine-tuning of the onsite energy of the added site results in a four-band model with three exactly flat bands for arbitrary hopping strength to the additional site. Second, we add a $s$-orbital at $1b$ Wyckoff position [yellow circle in Fig.~\ref{fig:fig1}(a)]. Note that we always consider only the subgroup $p2$ of the full point group $p6mm$ of the original kagome-3 model \cite{supp}. This additional orbital gives rise to an $A_{1b}$ band that does not remove the obstruction to an atomic limit. The non-trivial topology is now protected by $C_{2z}$ rather than $C_{2z}\mathcal{T}$ \cite{supp}. In this second case, it is not possible to achieve three exactly flat bands with finite range hopping. However, the addition of longer range hopping allows us to obtain three bands with $W/\delta\approx 0.03$ and a coupling strength to the addition site comparable to the first case. The different topological properties of these models can be read off the respective Wilson loop spectra \cite{Alexandradinata:2014} of Figs.~\ref{fig:fig1}(d)--(e): winding spectrum for the topological case, gapped for the trivial one. 

In the Monte Carlo simulations of these four-band models, we tune $\mu$ to achieve a filling $\nu=3/8$. This value corresponds to the half-filling of the lower three bands, where we intend to study the influence of fragile  topology on the superconducting behavior. The evolution of the superfluid weight as a function of temperature confirms the important role played by fragile topology, as can be seen in Fig.~\ref{fig:fig2}(c). In the topologically trivial model, $D_s(T)$ remains zero down to temperatures below the critical temperature of the original three-band model. On the other hand, the model with fragile topology protected by $C_{2z}$ symmetry behaves similarly to the kagome-3 model. 

\begin{figure}[t]
\includegraphics[]{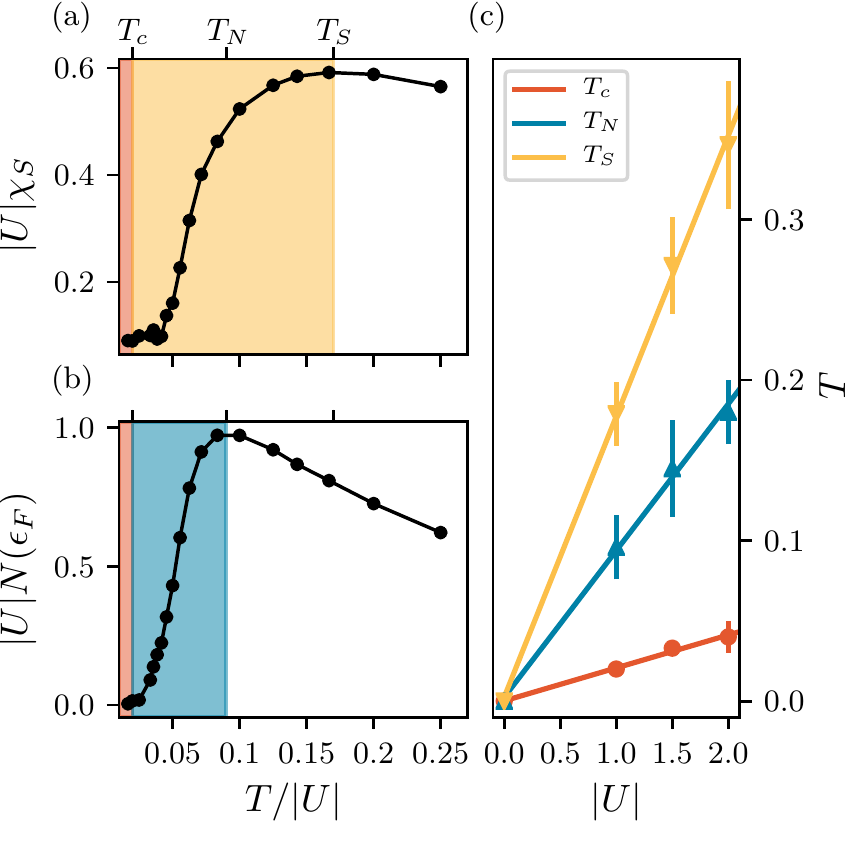}
\caption{\label{fig:fig3} (a) Spin susceptibility $\chi_{S}$ as a function of temperature $T$ for the Hubbard model on the kagome-3 lattice with $\lvert U \rvert =2$ and $L=6$. (b) Single-particle density of states $N(\epsilon_F)$ as a function of temperature $T$ for the same model. The shaded areas correspond to the superconducting state in dark orange, the range below $\chi_{S}$ peaks in yellow and that below the peak of $N(\epsilon_F)$ in blue. (c) Scaling of the critical temperatures $T_c$, $T_S$, and $T_N$ as a function of the interaction strength $\lvert U\rvert$. All these quantities show a linear scaling.}
\end{figure}

For completeness, we now turn our attention to the physics above the superconducting transition in the kagome-3 lattice. We first study the spin susceptibility
\begin{equation} 
\chi_{S}=\frac{1}{L^2}\int_0^\beta d\tau \langle S^z(\tau)S^z(0)\rangle,
\end{equation} 
with $S^z=\sum_i\left(c^\dagger_{i\uparrow}c_{i\uparrow}-c^\dagger_{i\downarrow}c_{i\downarrow}\right)$ \cite{supp}.
As shown in Fig.~\ref{fig:fig3}(a), it reaches a maximum at $T_S/\lvert U\rvert\approx 0.17$. 
This result points to the onset of singlet formation already above $T_c$ \cite{Randeria:1992}.

 Next, we investigate the single-particle density of states \cite{Trivedi:1995,supp} 
 \begin{equation}
 N(\epsilon_F)=\frac{\beta}{\pi L^2}\sum_{\alpha}\int_\text{BZ}d{\bf k}\,\langle c^{\phantom{\dagger}}_{\alpha\mathbf{k}}(\beta/2)c^\dagger_{\alpha\mathbf{k}}(0)\rangle,
 \end{equation}
where $\alpha$ is the sublattice index.
$N(\epsilon_F)$ peaks at temperature $T_N/\lvert U\rvert\approx  0.09$ and drops towards zero at lower temperatures, cf. Fig.~\ref{fig:fig3}(b). The temperature range $T_c < T < T_N$, where the opening of a gap reduces the density of states before the system turns superconducting, is associated with a pseudogap regime characterized by strong phase fluctuations \cite{Lee:2006}. Note that for $T<T_N$, also the spin susceptibility $\chi_{S}$ gets significantly suppressed.

Our results establish the importance of non-trivial fragile topology for the onset of superconductivity in flat bands. In summary, the signatures of a single-particle gap above the critical temperature are typical for attractive Hubbard models in the strong-coupling regime \cite{Trivedi:1995}. Moreover, the linear scaling with $\lvert U \rvert$ of the characteristic temperatures $T_c$, $T_S$, and $T_N$ shown in Fig.~\ref{fig:fig3}(c) is a generic feature of flat band physics for $\lvert U \rvert < \delta$ \cite{Volovik:2013,Hofmann:2019}. In particular, the pseudogap temperature scales linearly with $\lvert U \rvert$ regardless of whether it is identified with $T_S$ \cite{Randeria:1992} or $T_N$. Beyond these results, it was recently shown that band-topology can play a crucial role in the {\em strength} of superconductivity \cite{Hofmann:2019,Wang:2020}. While the authors of Ref.~[\onlinecite{Hofmann:2019}] considered the case of a Chern number, in our work the topological invariant ensuring a high critical temperature is a fragile one relevant for a broad class of time reversal invariant system. In particular, we prove how this new protecting mechanism is robust beyond the mean-field approximation of Ref.~[\onlinecite{Xie:2020}] but has important consequences for the fate of the superconducting state under the addition of trivial bands. Especially in two-dimensional systems, such additional bands naturally arise in tunnel-coupled heterostructures. This direct link between fragility and an observable quantity is an important step forward in our understanding of fragile topological insulators which have only a handful of known experimental signatures  \cite{Peri:2020,Song:2020,Lu:2020,Lian:2018,Liu:2019,Unal:2020}. Despite the infancy of this field, there is evidence that a myriad of materials and engineered structures possess this peculiar topology \cite{Song:2019, Alexandradinata:2019, Paz:2019, Wang:2019e}, calling for further studies of interactions in fragile bands \cite{Else:2019,Liu:2019,Latimer:2020}. 

\begin{acknowledgments}
	This work was supported by a grant from the Swiss National Supercomputing Centre (CSCS) under project ID eth5b. The auxiliary-field QMC simulations were carried out with the ALF package available at \href{https://alf.physik.uni-wuerzburg.de}{https://alf.physik.uni-wuerzburg.de}. V.P., and S.D.H. acknowledge support from the Swiss National Science Foundation, the NCCR QSIT, and the European Research Council under the Grant Agreement No. 771503 (TopMechMat). Z.S. and B.A.B. are supported by the Department of Energy Grant No. DE-SC0016239, the Schmidt Fund for Innovative Research, Simons Investigator Grant No. 404513, the Packard Foundation, the National Science Foundation EAGER Grant No. DMR-1643312, NSF-MRSEC No. DMR-1420541, BSF Israel US foundation No. 2018226, ONR No. N00014-20-1-2303, and the Princeton Global Network Funds.  
\end{acknowledgments}
%

\end{bibunit}
\let\addcontentsline\oldaddcontentsline

\onecolumngrid
\pagebreak
\setcounter{page}{1}
\thispagestyle{empty}
\begin{center}
	\textbf{\large Supplemental Material: Fragile topology and flat-band superconductivity in the \\ strong-coupling regime}\\[.2cm]
	
	  Valerio Peri,$^{1}$ Zhi-Da Song,$^{2}$ B. Andrei Bernevig,$^{2}$ and Sebastian D. Huber$^{1}$ \\[.1cm]
	  {\itshape ${}^1$Institute for Theoretical Physics, ETH Zurich, 8093 Z\"urich, Switzerland\\
	  ${}^2$Department of Physics, Princeton University, Princeton, New Jersey 08544, USA\\}
	(Dated: \today)\\[1cm]
	\end{center}
	\renewcommand{\thesection}{\Roman{section}}
	\begin{bibunit}[apsrev4-2]
	\renewcommand{\thetable}{S\arabic{table}}
\newcommand{\ph}{\phantom\dagger}
\renewcommand{\thefigure}{S\arabic{figure}}
\renewcommand{\theequation}{S\arabic{equation}}

\setcounter{figure}{0}
\setcounter{equation}{0}

\section{Tight-binding models}
\label{section:tb}
\subsection{Three-band model on the kagome-3 lattice}
\label{ssec:kagome3}
The kagome-3 lattice has basis vectors $\mathbf{a_1}=(1,0)$ and $\mathbf{a_2}=(1/2,\sqrt{3}/2)$. The reciprocal lattice vectors are $\mathbf{b_1}=(2\pi,-2\pi/\sqrt{3})$ and $\mathbf{b_2}=(0,4\pi/\sqrt{3})$. The unit cell contains three inequivalent sublattices $A$, $B$ and $C$. We set the distance among different sublattices in the same unit cell to unity. The tight-binding Hamiltonian defined on this lattice is:
\begin{equation}
	H_{\text{kin}}=\sum_{\mathbf{k},\sigma}c^\dagger_{\mathbf{k},\sigma}h(\mathbf{k})c^{\phantom{\dagger}}_{\mathbf{k},\sigma},
	\end{equation}
where $\sigma\in\{\uparrow,\downarrow\}$ is the electron's spin, $\mathbf{k}$ the electron's momentum and $c_{\sigma,\mathbf{k}}=\left(c_{\sigma,\mathbf{k}}^A,c_{\sigma,\mathbf{k}}^B,c_{\sigma,\mathbf{k}}^C\right)^T$ is the fermionic annihilation operator. The Bloch Hamiltonian $h(\mathbf{k})$ is given by:
\begin{equation}\resizebox{\columnwidth}{!}{$
	h(\mathbf{k})=
\begin{pmatrix}
	2\cos{\mathbf{k}\cdot\mathbf{a_2}} & 1 + e^{i \mathbf{k}\cdot\mathbf{a_1}}+e^{i \mathbf{k}\cdot\mathbf{a_2}} + e^{-i \mathbf{k}\cdot(\mathbf{a_1}-\mathbf{a_2})} & 1 + e^{i \mathbf{k}\cdot(\mathbf{a_1}-\mathbf{a_2})} + e^{i \mathbf{k}\cdot\mathbf{a_1}} + e^{-i \mathbf{k}\cdot\mathbf{a_2}} \\
	1 + e^{-i \mathbf{k}\cdot\mathbf{a_1}} + e^{-i \mathbf{k}\cdot\mathbf{a_2}} + e^{i \mathbf{k}\cdot(\mathbf{a_1}-\mathbf{a_2})} & 2\cos{\mathbf{k}\cdot(\mathbf{a_1}-\mathbf{a_2})}  & 1 + e^{-i \mathbf{k}\cdot\mathbf{a_2}}+e^{-i \mathbf{k}\cdot\mathbf{a_1}} + e^{i \mathbf{k}\cdot(\mathbf{a_1}-\mathbf{a_2})}\\
	1 + e^{-i \mathbf{k}\cdot(\mathbf{a_1}-\mathbf{a_2})}+e^{-i \mathbf{k}\cdot\mathbf{a_1}} + e^{i \mathbf{k}\cdot\mathbf{a_2}} & 1 + e^{i \mathbf{k}\cdot\mathbf{a_2}}+e^{i \mathbf{k}\cdot\mathbf{a_1}} + e^{-i \mathbf{k}\cdot(\mathbf{a_1}-\mathbf{a_2})} & 2\cos{\mathbf{k}\cdot\mathbf{a_1}} \\
\end{pmatrix}.$}
\end{equation}
Here, we choose a gauge where $h(\mathbf{k}+\mathbf{b_j})=h(\mathbf{k})$ for $j\in \{ 1,2\}$. We refer to this choice as the \emph{unit-cell gauge}. At times, it might be more convenient to use a gauge where $h(\mathbf{k}+\mathbf{b_j})=V(\mathbf{b_j})^{-1}h(\mathbf{k})V(\mathbf{b_j})$ for $j\in\{ 1,2\}$, with the unitary matrix $V(\mathbf{k})=\text{diag}[e^{-i \mathbf{k}\cdot\mathbf{a_2}/2},e^{-i \mathbf{k}\cdot(\mathbf{a_2}+\mathbf{a_1})/2},e^{-i \mathbf{k}\cdot\mathbf{a_1}/2}]$. We refer to this gauge choice as the \emph{periodic gauge}.
With the latter choice, also the periodic component of the Bloch wave function $u^n_{\mathbf{k}}$ transforms as $u^n_{\mathbf{k}+\mathbf{b_j}}=V(\mathbf{b_j})^{-1}u^n_{\mathbf{k}}$ for $j\in \{ 1,2\}$.

The spectrum of this model has two flat bands at energy $\epsilon=-2$ and one dispersive band $\epsilon=4+2[\cos{\mathbf{k}\cdot \mathbf{a_1}}+\cos{\mathbf{k}\cdot \mathbf{a_2}}+\cos{\mathbf{k}\cdot (\mathbf{a_1}-\mathbf{a_2})}]$. The smallest gap between the flat bands and the dispersive one is $\delta=3$ at $K=(2\pi/3,2\pi/\sqrt{3})$.

The full point group symmetry of this model is $p6mm$. The symmetry generators of this group are $C_{6z}$, $C_{3z}$, $C_{2z}$, $m_{10}$, and $m_{\bar{1}2}$. Here, $C_{nz}$ is a $\frac{2\pi}{n}$ rotation around an out-of-plane axis passing through $(0,0)$, and $m_{ij}$ is the mirror with respect to the plane perpendicular to the vector $i\mathbf{a_1} +j \mathbf{a_2}$, with $\mathbf{a_1}$ and $\mathbf{a_2}$ the basis vectors and $\bar{\imath}=-i$. The matrix representations of the symmetry generators are given for completeness together with their action in momentum space:
\begin{equation}
	C_6(\mathbf{k})=
	\begin{pmatrix}
		0 & 0 & 1 \\
		e^{-i \mathbf{k}\cdot(\mathbf{a}_1-\mathbf{a}_2)} & 0 & 0 \\
		0 & e^{i \mathbf{k}\cdot\mathbf{a}_2} & 0 
		\end{pmatrix}
	\qquad 
	\{k_x,k_y\} \rightarrow \left\{\frac{1}{2} k_x-\frac{\sqrt{3}}{2} k_y, \frac{\sqrt{3}}{2} k_x+\frac{1}{2}k_y\right\},
	\end{equation}
	
\begin{equation}
	C_3(\mathbf{k})=
	\begin{pmatrix}
		0 & e^{i \mathbf{k}\cdot\mathbf{a}_2} & 0 \\
		0 & 0 & e^{i \mathbf{k}\cdot\mathbf{a}_2} \\
		e^{i \mathbf{k}\cdot\mathbf{a}_2} & 0 & 0 
		\end{pmatrix}
	\qquad 
	\{k_x,k_y\} \rightarrow \left\{-\frac{1}{2} k_x-\frac{\sqrt{3}}{2} k_y, \frac{\sqrt{3}}{2} k_x-\frac{1}{2}k_y\right\},
	\end{equation}
	
\begin{equation}
	C_2(\mathbf{k})=
	\begin{pmatrix}
		e^{i \mathbf{k}\cdot\mathbf{a}_2} & 0 & 0 \\
		0 & e^{i \mathbf{k}\cdot(\mathbf{a}_1+\mathbf{a}_2)} & 0 \\
		0 & 0 & e^{i \mathbf{k}\cdot\mathbf{a}_1} 
		\end{pmatrix}
	\qquad 
	\{k_x,k_y\} \rightarrow \{-k_x,-k_y\},
	\end{equation}
	
\begin{equation}
	m_{10}(\mathbf{k})=
	\begin{pmatrix}
		0 & e^{i \mathbf{k}\cdot\mathbf{a}_1} & 0 \\
		e^{i \mathbf{k}\cdot\mathbf{a}_1} & 0 & 0 \\
		0 & 0 & e^{i \mathbf{k}\cdot\mathbf{a}_1}
		\end{pmatrix}
	\qquad 
	\{k_x,k_y\} \rightarrow \{-k_x,k_y\},
	\end{equation}
	
\begin{equation}
	m_{\bar{1}2}(\mathbf{k})=
	\begin{pmatrix}
		0 & e^{i \mathbf{k}\cdot\mathbf{a}_2} & 0 \\
		e^{-i \mathbf{k}\cdot(\mathbf{a}_1-\mathbf{a}_2)} & 0 & 0 \\
		0 & 0 & 1 
		\end{pmatrix}
	\qquad 
	\{k_x,k_y\} \rightarrow \{k_x,-k_y\}.
	\end{equation}
Note that the matrix representations depend on the chosen gauge. The representations given above refer to the unit-cell gauge and that is why they are momentum dependent. It is possible to choose different gauges such that these representations are momentum independent. The symmetry center is the  $1a$ Wyckoff position.

The $p6mm$ symmetry group is crucial to obtain doubly-degenerate exactly flat bands. Nevertheless, as discussed in Sec. \ref{ssec:symm}, it is not essential to protect their non-trivial topology. The topology of this model , as well as of all the others considered in our study, is entirely captured by $C_{2z}\mathcal{T}$ and $C_{2z}$. For this reason, we consider the kagome-3 model from the perspective of its subgroup $p2$, whose only generator is $C_{2z}$. In particular, both in the main text and the supplemental material we always refer to the maximal Wyckoff positions of the point group $p2$. 

\subsection{Trivial four-band model\label{ssec:trival4}}
We consider a tight-biding model with an additional $s$ orbital at the $1a$ Wyckoff position, i.e., the center of the kagome-3 unit cell. The additional site is coupled to all the nearest-neighbors sites. The Bloch Hamiltonian (in the unit-cell gauge) is:
\begin{equation}
	h_{4,\text{triv}}(\mathbf{k})=
\left(\begin{array}{c|ccc}
  t^2-2 & t(1 + e^{i\mathbf{k}\cdot\mathbf{a_2}}) & t (e^{i\mathbf{k}\cdot\mathbf{a_1}} + e^{i\mathbf{k}\cdot\mathbf{a_2}}) & t(1 + e^{i\mathbf{k}\cdot\mathbf{a_1}})  \\
\hline
  t(1 + e^{-i\mathbf{k}\cdot\mathbf{a_2}})&  \phantom{-} &  \phantom{-}  & \phantom{-}\\
  t (e^{-i\mathbf{k}\cdot\mathbf{a_1}} + e^{-i\mathbf{k}\cdot\mathbf{a_2}}) &  \phantom{-} & h(\mathbf{k}) &  \phantom{-}\\
  t(1 + e^{-i\mathbf{k}\cdot\mathbf{a_1}}) &  \phantom{-} &  \phantom{-} &  \phantom{-}\\
  \end{array}\right).
\end{equation}
The periodic gauge is obtained via the unitary transformation $V(\mathbf{k})=\text{diag}[1,e^{-i \mathbf{k}\cdot\mathbf{a_2}/2},e^{-i \mathbf{k}\cdot(\mathbf{a_2}+\mathbf{a_1})/2},e^{-i \mathbf{k}\cdot\mathbf{a_1}/2}]$.

\begin{figure}[t]
\includegraphics[]{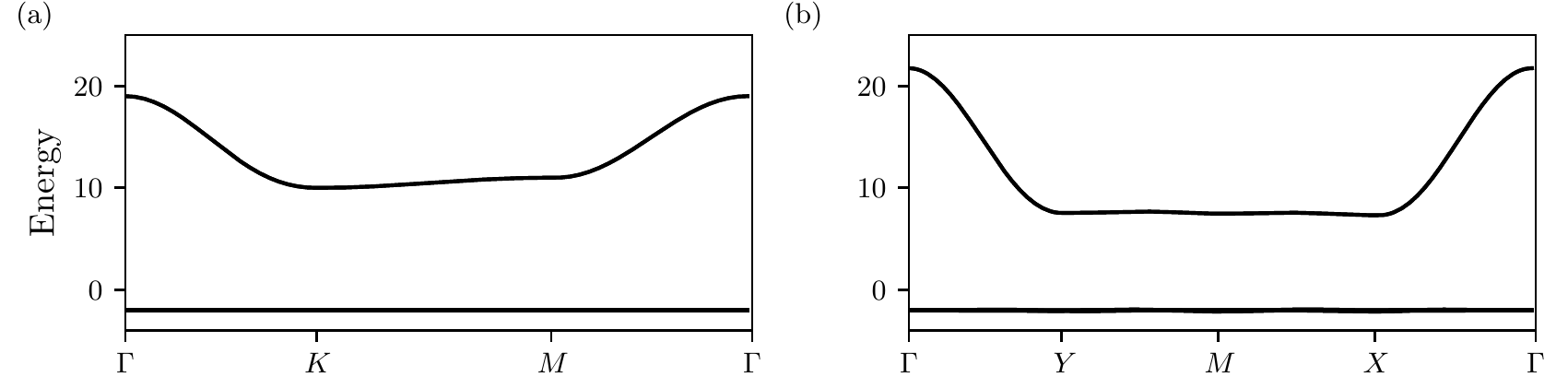}
\caption{\label{fig:fig1supp} (a) Spectrum along the high-symmetry line $\Gamma-K-M-\Gamma$ of the trivial tight-binding model with an additional $s$ orbital at the $1a$ Wyckoff position. The flat bands are triply degenerate. (b) Spectrum along the high-symmetry line $\Gamma-Y-M-X-\Gamma$ of the topological tight-binding model with an additional $s$ orbital at the $1b$ Wyckoff position. The nearly flat bands are again triply degenerate.}
\end{figure}

The additional band is exactly flat and degenerate with the two original flat bands of the kagome-3 lattice at $\epsilon=-2 \quad \forall t$, where $t$ is the hopping term between the original sites of the kagome-3 model and the additional site at the $1a$ Wyckoff position. In this study, we consider $t=3$ which leads to a gap $\delta=12$ at the $K$ point. The spectrum of this model is shown in Fig \ref{fig:fig1supp}(a). 

The full point group of this model is again $p6mm$. 

\subsection{Topological four-band model}
\label{ssec:topo4}
We can rather add an $s$ orbital at the $1b$ Wyckoff position, cf. Fig. 1(a) of the main text. A Bloch Hamiltonian (in the unit-cell gauge) describing this situation is:
\begin{equation}
	\label{eq:temp}
	h_{4,\text{topo}}(\mathbf{k})=
\left(\begin{array}{c|ccc}
  m & 0 & 0 & s\\
\hline
  0 &  \phantom{-} &  \phantom{-}  & \phantom{-}\\
  0 &  \phantom{-} & h(\mathbf{k}) &  \phantom{-}\\
  s &  \phantom{-} &  \phantom{-} &  \phantom{-}\\
  \end{array}\right).
\end{equation}
With this simple Hamiltonian, it is not possible to obtain three exactly flat bands in the spectrum of the model for arbitrary $m$ and $s$ values. Even allowing for more complicated nearest-neighbor hopping terms, we cannot obtain three exactly flat bands as in Sec. \ref{ssec:trival4}. To flatten the band, we then introduce longer range hopping terms. If $U_{\mathbf{k}}$ is the matrix that diagonalizes $h_{4,\text{topo}}(\mathbf{k})$, we can consider:
\begin{equation}
	h_{4,\text{topo}}(\mathbf{k})=U^{\phantom{\dagger}}_{\mathbf{k}} \Xi(\mathbf{k}) U^\dagger_{\mathbf{k}},
	\end{equation}
where $\Xi(\mathbf{k})$ is the diagonal matrix of the eigenvalues of $h_{4,\text{topo}}(\mathbf{k})$. We can replace $\Xi(\mathbf{k})$ with a new diagonal matrix $\tilde\Xi(\mathbf{k})$, where three entries are replaced with the desired flat bands energy: $\epsilon=-2$. A new Hamiltonian is then constructed as:
\begin{equation}
	\tilde h_{4,\text{topo}}(\mathbf{k}) = U^{\phantom{\dagger}}_{\mathbf{k}} \tilde\Xi(\mathbf{k}) U^\dagger_{\mathbf{k}}.
	\end{equation}
The real-space Hamiltonian is then retrieved by a simple Fourier transform. 
This procedure preserves the eigenvectors of the original Hamiltonian and its topology, while flattening the desired bands. The price to pay is the presence of hopping terms of arbitrary range.

With a careful optimization, we could achieve a set of three nearly flat bands with few additional hopping terms. In particular, we consider only couplings between sites at most two unit cells apart and with strength bigger than $0.02$. The resulting set of three flat bands is characterized by $W\approx 0.28$ and $\delta\approx 7.96$ which result in a flatness ratio $W/\delta \approx 0.03$. The additional site is coupled with a strength comparable to Sec. \ref{ssec:trival4}. The obtained spectrum is shown in Fig \ref{fig:fig1supp}(b). The unitary transformation to obtain the periodic gauge is $V(\mathbf{k})=\text{diag}[e^{-i \mathbf{k}\cdot\mathbf{a_1}/2},e^{-i \mathbf{k}\cdot\mathbf{a_2}/2},e^{-i \mathbf{k}\cdot(\mathbf{a_2}+\mathbf{a_1})/2},e^{-i \mathbf{k}\cdot\mathbf{a_1}/2}]$.

Note that this model reduces the point group from $p6mm$ to $p2mm$, still maintaining the $C_{2z}$ symmetry.

\section{Fragile topology}
\label{section:fragile}
\subsection{Wilson loops}
\label{ssec:wilson}
The nearly flat bands of magic-angle twisted bilayer graphene have fragile topology protected by $C_{2z} \mathcal T$. This symmetry is antiunitary and satisfies $\left(C_{2z} \mathcal T\right)^2=+1$. As we will outline below, a careful analysis identifies $C_{2z} \mathcal T$ as the protecting symmetry of the two flat bands of the kagome-3 lattice. Therefore, even when all crystalline symmetries are broken, our model possesses fragile topology in the presence of the composite $C_{2z} \mathcal T$ symmetry. 

To study the topological properties of the occupied bands, we use Wilson loop operators \cite{Alexandradinata:2014}. Indeed, the antiunitary nature of $C_{2z} \mathcal T$ prevents an analysis in terms of symmetry eigenvalues. Consider the Hamiltonian $h(\mathbf{k})$  and the eigenvectors of the two occupied flat bands in the periodic gauge: $u_{0\mathbf{k}}$ and $u_{1\mathbf{k}}$. We can construct an $m\times 2$ matrix of the occupied bands $U_\mathbf{k}=[u_{0\mathbf{k}},u_{1\mathbf{k}}]$, where $m$ is the number of inequivalent sublattices in the unit cell. In the periodic gauge, $h(\mathbf{k}+\mathbf{b_j})=V(\mathbf{b_j})^{-1}h(\mathbf{k})V(\mathbf{b_j})$ and $u^n_{\mathbf{k}+\mathbf{b_j}}=V(\mathbf{b_j})^{-1}u^n_{\mathbf{k}}$ for $j\in\{ 1,2\}$. We consider Wilson loops along the reciprocal lattice vectors and parametrize $\mathbf{k}=k_1\mathbf{b_1}+k_2\mathbf{b_2}$, with $k_j \in [0,1]$. The discretized Wilson loop integrated along $k_2$ is then defined as \cite{Wilczek:1984,Zak:1989,Alexandradinata:2014}:
\begin{equation}
	\label{eq:wl}
	\mathcal{W}(k_1)=U^\dagger_{k_1,0}U^{\phantom{\dagger}}_{k_1,\frac{1}{N}}U^\dagger_{k_1,\frac{1}{N}}U^{\phantom{\dagger}}_{k_1,\frac{2}{N}} \dots U^\dagger_{k_1,\frac{N-1}{N}}V(\mathbf{b_2})U_{k_1, 0},
	\end{equation}
where $N$ is the sample's size and is assumed to be large. However, a finite albeit large $N$ renders Eq.~\eqref{eq:wl} non-unitary. We ensure unitarity via the singular value decomposition $$G_{k_1,\frac{j}{N}}=U^\dagger_{k_1,\frac{j}{N}}U^{\phantom{\dagger}}_{k_1,\frac{j+1}{N}}=SDP^\dagger,$$where $D$ is a diagonal matrix. We then define the Wilson loop operator in terms of the matrix $F_{k_1,\frac{j}{N}}=SP^\dagger$:
\begin{equation}
	\label{eq:Wunit}
	\mathcal{W}(k_1)=F_{k_1,0}F_{k_1,\frac{1}{N}} \dots F_{k_1,\frac{N-1}{N}}V(\mathbf{b_2})F_{k_1, 0}.
	\end{equation}
The discretized Wilson loop of Eq.\eqref{eq:Wunit} is a unitary operator which can be expressed as $\mathcal{W}(k_1)=e^{i\mathcal{H}_\mathcal{W}(k_1)}$. Here, the Wilson Hamiltonian $\mathcal{H}_\mathcal{W}(k_1)$ is defined as:
\begin{equation}
	\mathcal{H}_\mathcal{W}(k_1)= - i\,\text{log}\mathcal{W}(k_1).
	\end{equation}
In the kagome-3 lattice, $\mathcal{H}_\mathcal{W}(k_1)$ for the two flat bands is a $2\times 2$ Hermitian matrix whose eigenvalues form the Wilson spectrum shown in Fig. 1(c) of the main text. 

The Wilson loop operators can be readily computed also for the four-band models. There, we investigate the topology of the lower three bands of the model. Therefore, we have $4\times 3$ $U_\mathbf{k}$ matrices and the Wilson loop Hamiltonian is a $3 \times 3$ Hermitian matrix. The Wilson spectra of these models are shown in Figs. 1(d)--(e) of the main text.

It is possible to assess the topology protected by $C_{2z} \mathcal T$ directly from the Wilson spectrum \cite{Song:2019a,Xie:2020,Ahn:2019,Bouhon:2018,Bouhon:2020,Ahn:2018}. $C_{2z} \mathcal T$ introduces a topological $\mathbb{Z}$ classification for two occupied bands and in the following we briefly review how it comes about with the help of Wilson loop operators. The $C_{2z} \mathcal T$ action in momentum space is local. Hence, it is possible to find a gauge where this symmetry is represented by the identity matrix. Such gauge, for the kagome-3 model, corresponds to the periodic gauge defined in Sec. \ref{ssec:trival4}. In this case, $(C_{2z} \mathcal T)^{-1}h(\mathbf{k})(C_{2z} \mathcal T)=h(\mathbf{k})=h^*(\mathbf{k})$ and the Bloch Hamiltonian is real. The same condition applies to the periodic component of the Bloch wave function $u^{\phantom{*}}_\mathbf{k}=u^*_\mathbf{k}$. 
Since the periodic Bloch wave functions are real, also $\mathcal{W}(k_1)$ is real [see Eq.~\eqref{eq:wl}]:
\begin{equation}
	\label{eq:realW}
	\mathcal{W}(k_1)=\mathcal{W}^*(k_1).
\end{equation}
The reality condition on the Wilson loop imposes an effective particle-hole symmetry on $\mathcal{H}_\mathcal{W}(k_1)$:
\begin{equation}
	\label{eq:phhw}
	\mathcal{C}^{-1}\mathcal{H}_\mathcal{W}(k_1)\mathcal{C}=\mathcal{H}^*_\mathcal{W}(k_1)=-\mathcal{H}_\mathcal{W}(k_1) \quad \text{mod}\, 2,
	\end{equation}
where the particle-hole conjugation is simply represented by complex conjugation: $\mathcal{C}=K$. 
A $2\times 2$ Hermitian matrix can always be expressed in the Pauli matrices basis as:
\begin{equation}
	\label{eq:hPauli}
	\mathcal{H}_\mathcal{W}(k_1)=d_0(k_1)\sigma^0+d_1(k_1)\sigma^1+d_2(k_1)\sigma^2+d_3(k_1)\sigma^3.
	\end{equation}
To satisfy Eq.~\eqref{eq:phhw}, the only non-zero component is $d_2(k_1)$. This result is the important restriction imposed by $C_{2z} \mathcal T$ on the Wilson Hamiltonian. One then considers the winding number of $d_2(k_1)$ as a function of $k_1$ \cite{Song:2019a}:
\begin{equation}
	\label{eq:1DWind}
	w=\int_0^{2\pi} \frac{\text{d}k_1}{2\pi}\frac{\partial_{k_1}d_2(k_1)}{d_2(k_1)}.
	\end{equation}
This quantity introduces a $C_{2z} \mathcal T$-protected $\mathbb{Z}$ index for a two-band model associated to the first homotopy of the circle: $\pi_1(S^1)=\mathbb{Z}$. This index corresponds to the Euler class $e_2$ of oriented real vector bundles. Note that it applies exclusively to two occupied bands as the decomposition of Eq.~\eqref{eq:hPauli} does not hold otherwise. The required restriction on the number of occupied bands points to the fragile nature of the $\mathbb{Z}$ index protected by $C_{2z} \mathcal T$.
The winding number of the two flat bands of the kagome-3 model confirms that they have $e_2=1$, protected by $C_{2z} \mathcal T$ symmetry. 


The Wilson loop spectrum also allows to asses the topology protected by the crystalline symmetry $C_{2z}$ \cite{Alexandradinata:2014}. $C_{2z}$ defines a $\mathbb{Z}_2$ index that corresponds to the parity of the winding of the Wilson loop spectrum. The winding of the Wilson loop spectrum in Fig. 1(e) of the main text confirms that the model of Sec. \ref{ssec:topo4} is topologically non-trivial. On the other hand, the model of Sec. \ref{ssec:trival4} is trivial, cf. Fig. 1(d) of the main text. We will further address the nature of this model's topology in Sec \ref{ssec:symm} from a symmetry indicators perspective. 

We further stress that the absence of winding in the Wilson loop spectrum of  the model of Sec. \ref{ssec:trival4} highlights the fragile nature of the topology in the kagome-3's flat band. Namely, the addition of a trivial Bloch band resolves the topological obstruction to an atomic limit.

\subsection{Symmetry eigenvalues}
\label{ssec:symm}
In the presence of more than two bands, the Euler class is not well defined. The relevant characteristic class is the second Stiefel-Whitney class $w_2$. The latter can be computed directly from the crossing of the Wilson loop spectrum at $\pi$ \cite{Ahn:2019a}. For all of the tight binding models considered in this work we find $\lvert w_2 \rvert=1$. Nevertheless, $w_2$ is not related to an obstruction to symmetric localized Wannier functions but rather distinguishes among inequivalent atomic insulators. As such, it is not linked to a bound on the Wannier functions' localization which favors superconducting instabilities at the mean-field level. 
To characterize the topological properties of the three lower bands of the model of Secs. \ref{ssec:trival4} and \ref{ssec:topo4}, we rather restore to an analysis of the $C_{2z}$ symmetry eigenvalues in the framework of Topological Quantum Chemistry \cite{Bradlyn:2017}. 

Let us first discuss the kagome-3 model of Sec. \ref{ssec:kagome3}. Its point group is $p6mm$. To capture its topological properties, however, it suffices to consider the point group $p2$, whose generator is $C_{2z}$. Its matrix representation and action in momentum space are:
%
	
\begin{equation}
	\label{eq:c2matrix}
	C_{2z}(\mathbf{k})=\text{diag}\left(e^{i \mathbf{k}\cdot\mathbf{a}_2},e^{i \mathbf{k}\cdot(\mathbf{a}_1+\mathbf{a}_2)},e^{i \mathbf{k}\cdot\mathbf{a}_1}\right)
	\qquad 
	\{k_x,k_y\} \rightarrow \{-k_x,-k_y\}.
	\end{equation}
	
%
Note that the matrix representation depends on the chosen gauge. The one of Eq. \eqref{eq:c2matrix} refers to the unit-cell gauge and is momentum dependent. The out-of-plane rotation axis passes through the $1a$ Wyckoff position, cf. Fig. 1(a) of the main text.

We can now look at the symmetry eigenvalues at the high-symmetry points in momentum space, i.e., $\Gamma=(0,0)$, $X=(\pi,-\pi/\sqrt{3})$, $Y=(0,2\pi/\sqrt{3})$, and $M=(\pi,\pi/\sqrt{3})$, to determine the irreducible representations. The little group at these points is $p2$, which has two irreducible representations: $K_1$ and $K_2$, respectively even and odd under $C_{2z}$. Here $K$ stands for an arbitrary high-symmetry-point, e.g., $K_1\rightarrow \Gamma_1$ at the $\Gamma$ point. 
The lowest two flat bands realize the irreducible representations $2\Gamma_1$, $2X_2$, $2Y_2$ and $2M_2$. The dispersive band $\Gamma_1$, $X_1$, $Y_1$, and $M_1$. 

We can compare the representations at high-symmetry points with those of the elementary band representations (EBRs) of $p2$. The latter correspond to all the bands that can be realized by localized symmetric Wannier functions. Via this analysis, we establish the following EBRs decomposition:
\begin{equation}
	\label{eq:fragiledecoc2}
		\begin{split}
		\text{flat bands}:&\quad (A)_{1b}\oplus(A)_{1c}\oplus(A)_{1d}\ominus(A)_{1a} ,\\
		\text{dispersive band}:& \quad (A)_{1a}.
		\end{split}
\end{equation} 
This result establishes the fragile topological nature of the two lowest flat bands from a crystalline perspective. Indeed, they cannot be expressed as a direct sum of EBRs with positive coefficients but such a decomposition is possible after the addition of a trivial band, e.g., $(A_1)_{1a}$. The table of the EBRs of $p2$ is here reported for completeness in Tab. \ref{tab:p2EBR} \cite{Bilbao}. We stress again that the fragile topology of the kagome-3 model survives also under perturbations that break $C_{2z}$, as long as $C_{2z} \mathcal T$ is preserved, cf. Sec \ref{ssec:wilson}.

\begin{table}[!ht]
  \begin{tabular}{lcccccccc}
    \hline
    \hline
  & $(A)_{1a}$ & $(B)_{1a}$ & $(A)_{1b}$ & $(B)_{1b}$ & $(A)_{1c}$
  & $(B)_{1c}$ & $(A)_{1d}$ & $(B)_{1d}$  \\
  \hline
  $\Gamma$ & $\Gamma_1$ & $\Gamma_2$ & $\Gamma_1$ & $\Gamma_2$ & $\Gamma_1$ & $\Gamma_2$ & 
  $\Gamma_1$ & $\Gamma_2$ 
  \\
  $X$ & $X_1$ & $X_2$ & $X_1$ & $X_2$ & $X_2$ & $X_1$ & 
  $X_2$ & $X_1$ 
  \\
  $Y$ & $Y_1$ & $Y_2$ & $Y_2$ & $Y_1$ & $Y_1$ & $Y2$ & 
  $Y_2$ & $Y_1$  
  \\
  $M$ & $M_1$ & $M_2$ & $M_2$ & $M_1$ & $M_2$ & $M_1$ & 
  $M_1$ & $M_2$\\
  \hline
  \hline
  \end{tabular}
  \caption{{\bf Elementary band representations for $\mathbf{p2}$.} \cite{Bilbao} The upper row indicates the name of the EBR as $(K)_\ell$, where $K$ is the irreducible representation of the orbital that induces the EBR and $\ell$ its maximal Wyckoff position. Each column contains the irreducible representations at high symmetry points in momentum space for the different EBRs.}

  \label{tab:p2EBR}

  \end{table}
  
The $C_{2z}$ representation is readily adapted to our four-band models. For the model of Sec. \ref{ssec:trival4}, we have: 
\begin{equation}
	C_{2z}(\mathbf{k})=\text{diag}\left(1,e^{i \mathbf{k}\cdot\mathbf{a}_2},e^{i \mathbf{k}\cdot(\mathbf{a}_1+\mathbf{a}_2)},e^{i \mathbf{k}\cdot\mathbf{a}_1}\right).
	\end{equation}
For the model of Sec. \ref{ssec:topo4}, instead,
\begin{equation}
	C_{2z}(\mathbf{k})=\text{diag}\left(e^{i \mathbf{k}\cdot\mathbf{a}_1},e^{i \mathbf{k}\cdot\mathbf{a}_2},e^{i \mathbf{k}\cdot(\mathbf{a}_1+\mathbf{a}_2)},e^{i \mathbf{k}\cdot\mathbf{a}_1}\right).
	\end{equation}
	
	In the first case, we have an additional $(A)_{1a}$ EBR. The decomposition of the lowest three bands is $(A)_{1b}\oplus(A)_{1c}\oplus(A)_{1d}$ and they are topologically trivial. In the second instance, the additional orbital realizes the $(A)_{1b}$ EBR. The EBRs decomposition of the three lowest bands $2(A)_{1b}\oplus(A)_{1c}\oplus(A)_{1d}\ominus(A)_{1a} $ establishes their fragile topology protected by $C_{2z}$.  

\section{Superfluid weight in the mean-field approximation}
\label{section:ds}
\subsection{Uniform pairing condition}
\label{ssec:uniform}
The BCS wave function is an exact ground state at zero temperature for the attractive Hubbard model on bipartite lattices \cite{Peotta:2015,Tovmasyan:2016}. The situation is more complicate for non-bipartite lattices as the kagome-3. Here, we present the uniform pairing condition under which the BCS wave function is an exact zero-temperature ground state for these lattices. For a careful derivation of the results reviewed here see Ref.~[\onlinecite{Tovmasyan:2016}].  

Consider the matrix of the occupied bands $U_\mathbf{k}=[u_{0\mathbf{k}},u_{1\mathbf{k}}]$, where $u_{0\mathbf{k}}$ and $u_{1\mathbf{k}}$ are the Bloch wave functions of the two flat bands of the kagome-3 lattice. One constructs the matrix
\begin{equation}
	\label{eq:uniformPair}
	P=\frac{V_c}{(2\pi)^2}\int_\text{BZ} \text{d}^2k\,  U^{\phantom{\dagger}}_\mathbf{k}U^\dagger_\mathbf{k},
\end{equation}
where $V_c$ is the area of the two-dimensional unit cell.
The uniform pairing condition requires that all the diagonal entries of the matrix $P$ are identical (if non-zero) \cite{Tovmasyan:2016}. We can explicitly verify that the flat bands of the kagome-3 lattice satisfy this condition. The three lowest bands of the four-band models, on the other hand, do not satisfy it.  


The uniform pairing condition of the two flat bands of the kagome-3 lattice justifies the use of BCS theory to estimate the superfluid weight at zero temperature. We stress that the use of BCS theory is otherwise not justified at finite temperature and in the strong-coupling regime. Moreover, the uniform pairing condition for the validity of the BCS approximation is derived in the isolated and exactly flat band limit. The presence of a finite gap $\lvert U \rvert/\delta\neq 0$ might invalidate its conclusions.

\subsection{Zero temperature mean-field superfluid weight}
In this section, we briefly review how non-trivial fragile topology bounds the superfluid weight in the BCS mean-field approximation. Refs. [\onlinecite{Peotta:2015}] and [\onlinecite{Xie:2020}] offer a detailed derivation of the results presented here.  

Consider $N$ occupied bands with eigenstates $u_{j\mathbf{k}}$ for $j\in[1,N]$ and the matrix $U_\mathbf{k}=[u_{1\mathbf{k}},u_{2\mathbf{k}},\dots,u_{N\mathbf{k}}]$. The quantum geometric tensor (QGT) $\theta$ is defined as \cite{Provost:1980,Xie:2020,Neupert:2013,Peotta:2015}:
\begin{equation}	
	\theta_{ij}=\partial_{k_i}U^\dagger(\mathbf{k})\left[\mathbb{I}-U(\mathbf{k})U^\dagger(\mathbf{k})\right]\partial_{k_j}U(\mathbf{k}).
\end{equation}
The real and imaginary part of $\theta$ are respectively
\begin{equation}
	\label{eq:fubiniStudy}
	\resizebox{\columnwidth}{!}{$
	\begin{split}
	\text{Re}[\theta_{ij}]&=\mathfrak{g}_{ij}=\frac{1}{2}\left[\theta_{ij}+\theta^\dagger_{ij}\right]\\
	&= \frac{1}{2}\left[\partial_{k_i}U^\dagger(\mathbf{k})\partial_{k_j}U(\mathbf{k})+\partial_{k_j}U^\dagger(\mathbf{k})\partial_{k_i}U(\mathbf{k})+U^\dagger(\mathbf{k})\partial_{k_i}U(\mathbf{k})U^\dagger(\mathbf{k})\partial_{k_j}U(\mathbf{k})+U^\dagger(\mathbf{k})\partial_{k_j}U(\mathbf{k})U^\dagger(\mathbf{k})\partial_{k_i}U(\mathbf{k})\right],
	\end{split}$}
\end{equation}
\begin{equation}
	\label{eq:berryCurvature}
	\resizebox{\columnwidth}{!}{$
	\begin{split}
	\text{Im}[\theta_{ij}]=&\frac{1}{2i}\left[\theta_{ij}-\theta^\dagger_{ij}\right]\\=
	&
	\frac{1}{2i}\left[\partial_{k_i}U^\dagger(\mathbf{k})\partial_{k_j}U(\mathbf{k})-\partial_{k_j}U^\dagger(\mathbf{k})\partial_{k_i}U(\mathbf{k})+U^\dagger(\mathbf{k})\partial_{k_i}U(\mathbf{k})U^\dagger(\mathbf{k})\partial_{k_j}U(\mathbf{k})-U^\dagger(\mathbf{k})\partial_{k_j}U(\mathbf{k})U^\dagger(\mathbf{k})\partial_{k_i}U(\mathbf{k})\right].
	\end{split}$}
\end{equation}
Note that $\text{Re}[\theta_{ij}]=\text{Re}[\theta_{ji}]$ while $\text{Im}[\theta_{ij}]=-\text{Im}[\theta_{ji}]$. The Fubini-Study metric is defined as $g_{ij}=\text{Tr}\,\mathfrak{g}_{ij}$ and induces a metric on the complex projective Hilbert space. On the other hand, the Berry connection, defined as $\mathbf{A}=iU^\dagger(\mathbf{k})\partial_{\mathbf{k}}U(\mathbf{k})$, and the Berry curvature, $\mathcal{F}_{ij}=\partial_{k_i}A_j-\partial_{k_j}A_i-i[A_i,A_j]$ , are directly related to Eq. \eqref{eq:berryCurvature}: 
\begin{equation}
	\text{Im}[\theta_{ij}]=-\frac{1}{2}\mathcal{F}_{ij}.
\end{equation}
Therefore, the QGT relates the Fubini-Study metric and the Berry curvature:
\begin{equation}
	\theta_{ij}=\mathfrak{g}_{ij}-\frac{i}{2}\mathcal{F}_{ij}.
\end{equation}
From the properties of the QGT, $\mathfrak{g}_{ij}=\mathfrak{g}_{ji}$ and $\mathcal{F}_{ij}=-\mathcal{F}_{ji}$ follow. Moreover, the fact that $\theta$ is positive definite allows formulating lower bounds on the Fubini-Study metric in terms of the Berry curvature.  


In the BCS approximation at zero temperature, the superfluid weight of exactly flat bands can be computed in terms of the Fubini-Study metric \cite{Peotta:2015}:
\begin{equation}
	\label{eq:dsgij}
	D_{s,ij}(T=0)=\frac{\Delta \sqrt{\nu_F(1-\nu_F)}}{V_{BZ}}\int_{\text{BZ}}\text{d}^2 k\, g_{ij}(\mathbf{k}),
\end{equation}
where $\nu_F$ is the filling factor of the flat bands considered, $\Delta$ is the BCS superconducting gap and $V_{BZ}$ is the area of the two-dimensional Brillouin zone. To compare the zero-temperature mean-field prediction to our simulation results, we need to solve the gap equation in terms of the local Hubbard interaction $\lvert U \rvert$. This task can be carried out explicitly in the limit $\delta \gg \lvert U \rvert \gg W$ and when the uniform pairing condition is satisfied \cite{Peotta:2015}. The result is $\Delta=\lvert U\rvert n_\phi\sqrt{\nu_F(1-\nu_F)}$ \cite{Peotta:2015}, where $n_\phi=1/3$ is the inverse of the number of sublattices over which the flat bands eigenstates have non-zero weight.

The QGT tensor relates the Fubini-Study metric to the Berry curvature of the occupied bands. Taking advantage of the positive definitive nature of the QGT, the integral of the Fubini-Study metric can be bounded from below by the bands's topological invariants. In particular, such bound was proved in Ref. [\onlinecite{Xie:2020}] for two bands with non-zero Euler class $e_2$:
\begin{equation}
	\label{eq:xieBound}
	D_s(T=0)\geq\frac{2\pi\lvert U \rvert n_\phi\nu_F(1-\nu_F)}{V_{BZ}}\lvert e_2\rvert.
\end{equation}

Note that Eqs.~\eqref{eq:dsgij} and \eqref{eq:xieBound} require the uniform pairing condition of Sec. \ref{ssec:uniform} and hold exclusively at zero temperature in the mean-field approximation. We can compare the value obtained from these expressions to our Monte Carlo simulations. The two flat bands of the kagome-3 lattice have $\lvert e_2 \rvert=1$. In turn, the mean-field value for the superfluid weight $D_s$ is obtained:
\begin{equation}
	D_s=\frac{\lvert U \rvert n_\phi\nu_F(1-\nu_F)}{2V_{BZ}}\int_{\text{BZ}}\text{d}^2 k\,\text{Tr}\, g(\mathbf{k})\approx 0.029 > 0.023 \approx \frac{2\pi\lvert U \rvert n_\phi\nu_F(1-\nu_F)}{V_{BZ}}\lvert e_2\rvert.
\end{equation}
Here, we used $\lvert U \rvert=2$, $n_\phi=1/3$, $V_{BZ}=8\pi^2/\sqrt{3}$ and a filling $\nu_F=1/2$, which corresponds to a system's filling $\nu=1/3$. As shown in Fig. 2(a) of the main text, the superfluid weight obtained from Monte Carlo simulations saturates the mean-field topological bound at low temperatures. 

\subsection{Symmetry constraints on the superfluid weight}
\label{ssec:symmDs}
We consider the superfluid weight along the orthogonal basis $(\mathbf{e_x},\mathbf{e_y})$, with $\mathbf{e_x}=(1,0)$ and $\mathbf{e_y}=(0,1)$:
\begin{equation}
	D_s=\begin{pmatrix}
	 D_{s,xx} & D_{s,xy} \\
	 D_{s,yx} & D_{s,yy}
	\end{pmatrix}.
\end{equation}
Since the Fubini-Study metric of Eq. \eqref{eq:fubiniStudy} is symmetric, the geometric contribution of Eq. \eqref{eq:dsgij} to the superfluid weight satisfies $D_{s,xy}=D_{s,yx}$. $C_{6z}$ symmetry requires invariance of $D_s$ under $\pi/3$ rotations: $R_{\pi/3}^TD_sR_{\pi/3}=D_s$, and 
\begin{equation}
	R_{\pi/3}=\begin{pmatrix}
	1/2 & \sqrt{3}/2\\
	-\sqrt{3}/2 & 1/2
	\end{pmatrix}.
\end{equation}
This condition imposes $D_{s,xy}=0$ and $D_{s,xx}=D_{s,yy}$ and it applies to the models of Secs. \ref{ssec:kagome3} and \ref{ssec:trival4}. In this work, we always consider a gauge potential $\mathbf{A}=A\mathbf{\hat{x}}$ and identify $D_{s,xx}$=$D_s$.

\section{Auxiliary-field quantum Monte Carlo}
The auxillary-field quantum Monte Carlo (QMC) simulations of this work were carried out with the ALF package available at \href{https://alf.physik.uni-wuerzburg.de}{https://alf.physik.uni-wuerzburg.de}. In Ref.~[\onlinecite{Bercx:2017}], one finds a discussion of the technical details of the code. For completeness, here we review the main ideas behind the auxiliary-field QMC method. We do not discuss further subtleties in the Monte Carlo updates and sampling techniques that can be found in Ref.~[\onlinecite{Bercx:2017}] and references therein. For example, the stabilization of the Green's functions or the Jackknife resampling method.  

An infamous obstacle to fermionic Monte Carlo simulations is the sign problem associated to the fermionic statistic. Nevertheless, there are classes of model that can be efficiently simulated without incurring into this problem \cite{Wu:2005,Li:2016,Li:2019}. For example, the attractive Hubbard model $H=H_K+H_I$, where
\begin{equation}
	H_K=\sum_{ij,\sigma}t_{ij,\sigma}\left(c^\dagger_{i\sigma}c^{\ph}_{j\sigma}+c^\dagger_{j\sigma}c^{\ph}_{i\sigma}\right)-\mu\sum_{i,\sigma}n_{i\sigma},
\end{equation}
and 
\begin{equation}
	H_I=U\sum_i n_{i\uparrow}n_{i\downarrow}.
\end{equation}
Here, $\mu$ is the chemical potential, $U<0$ the effective local Hubbard interaction, $t_{ij}$ the hopping parameter, and $c_{i\sigma}$ the fermionic annihilation operator at site $i$ with spin $\sigma$. Finally, we define $n_{i\sigma}=c^\dagger_{i\sigma}c^{\ph}_{i\sigma}$ and $n_i=\sum_{\sigma}n_{i\sigma}$.

A first important step to illustrate how the sign problem does not affect this model is to decouple the fermionic interaction via a Hubbard-Stratonovich transformation \cite{Hubbard:1959,Hirsch:1983}. This transformation is based on the identity:
\begin{equation}
	e^{\frac{1}{2}A^2}=\sqrt{2\pi}\int\text{d}xe^{-\frac{1}{2}x^2-Ax}.
\end{equation} 
We hence manipulate $H_I$ to express it as a square of operators. This task can be performed in multiple ways. We choose to decouple in the density channel in order to preserve the SU(2) symmetry of the original model \cite{Hirsch:1983}. One observes that
\begin{equation}
	\begin{split}
	H'_I&=\frac{U}{2}\sum_i\left[\sum_\sigma\left(c^\dagger_{i\sigma}c^{\ph}_{i\sigma}-\frac{1}{2}\right)\right]^2\\
	&=\frac{U}{2}\sum_i\left(n_{i\uparrow}+n_{i\downarrow}-1\right)^2\\
	&=U\sum_i n_{i\uparrow}n_{i\downarrow}-\frac{U}{2}\sum_i\left(n_i-1\right),
	\end{split}
\end{equation}
where we used $n_{i\sigma}^2=n^{\phantom{2}}_{i\sigma}$. $H_I$ and $H'_I$ differ only by a renormalization of the chemical potential and a global energy shift. 
From now on, we focus on the Hamiltonian:
\begin{equation}
	\label{eq:wholeH}
	H=\sum_{ij,\sigma}t_{ij,\sigma}\left(c^\dagger_{i\sigma}c^{\ph}_{j\sigma}+c^\dagger_{j\sigma}c^{\ph}_{i\sigma}\right)-\mu\sum_{i,\sigma}n_{i\sigma}+\frac{U}{2}\sum_i\left[\sum_\sigma\left(c^\dagger_{i\sigma}c^{\ph}_{i\sigma}-\frac{1}{2}\right)\right]^2.
\end{equation}

The goal is to obtain a simple expression for the partition function to perform Monte Carlo sampling:
\begin{equation}
	Z=\text{Tr}\,\left(e^{-\beta H}\right).
\end{equation} 
The first step is the discretization of imaginary time $\beta=\Delta\tau L_\tau$ and the Trotter decomposition that introduces a systematic error of order $\mathcal{O}(\Delta\tau^2)$. Secondly, we introduce auxiliary fields in order to decouple the quartic interaction term via Hubbard-Stratonovich (HS) transformation. In particular, The ALF package implementation performs a discrete HS transformation with the help of two fields $\gamma$ and $\eta$ \cite{Bercx:2017}:
\begin{equation}
	e^{-\Delta\tau\frac{U}{2}\left[\sum_\sigma\left(c^\dagger_{i\sigma}c^{\ph}_{i\sigma}-\frac{1}{2}\right)\right]^2}=\sum_{l=\pm 1,\pm2}\gamma(l)e^{\sqrt{-\Delta\tau\frac{U}{2}}\eta(l)\sum_\sigma\left(c^\dagger_{i\sigma}c^{\ph}_{i\sigma}-\frac{1}{2}\right)}+\mathcal{O}(\Delta\tau^4).
\end{equation}
Note that thanks to the Trotter error of order $\mathcal{O}(\Delta\tau^2)$, the discrete HS transformation is nearly exact. Here, the fields take the following values:
\begin{equation}
	\begin{split}
		\gamma(\pm 1)=1+\frac{\sqrt{6}}{3},\qquad & \eta(\pm 1)=\pm\sqrt{2\left(3-\sqrt{6}\right)}, \\
		\gamma(\pm 2)=1-\frac{\sqrt{6}}{3},\qquad &  \eta(\pm 2)=\pm\sqrt{2\left(3+\sqrt{6}\right)}.
	\end{split}
\end{equation}
The goal is to replace the sampling over the fermionic configurations with the sampling over the auxiliary fields configurations $C$:
\begin{equation}
	Z=\text{Tr}\,\left(e^{-\beta H}\right)=\sum_{C}e^{-S(C)}+\mathcal{O}(\Delta\tau^2),
\end{equation}
where $S(C)$ is the action of non-interacting fermions and the auxiliary fields in the configuration $C$. Note that this action is a priori not real and there is no justification to treat $e^{-S(C)}$ as a probability weight. 

After the HS transformation, the partition function $Z$ is expressed as: 
\begin{equation}
	\begin{split}
	Z&=\text{Tr}\,\left(e^{-\beta H}\right)\\
	&=\text{Tr}\,\left[\prod_\tau e^{-\Delta\tau\frac{U}{2}\sum_i\left[\sum_\sigma\left(c^\dagger_{i\sigma}c^{\ph}_{i\sigma}-\frac{1}{2}\right)\right]^2} e^{-\Delta\tau\left[\sum_{ij,\sigma}t_{ij,\sigma}\left(c^\dagger_{i\sigma}c^{\ph}_{j\sigma}+c^\dagger_{j\sigma}c^{\ph}_{i\sigma}\right)-\mu\sum_{i,\sigma}n_{i\sigma}\right]}\right]+\mathcal{O}(\Delta\tau^2)\\
	&=\sum_{C}\left(\prod_\tau\prod_i \gamma_{i,\tau}\right)
	\text{Tr}\left[\prod_\tau\left( e^{\sqrt{-\Delta\tau\frac{U}{2}}\sum_{i,\sigma}\eta_{i,\tau}\left(c^\dagger_{i\sigma}c^{\ph}_{i\sigma}-\frac{1}{2}\right)} e^{-\Delta\tau\left[\sum_{ij,\sigma}t_{ij,\sigma}\left(c^\dagger_{i\sigma}c^{\ph}_{j\sigma}+c^\dagger_{j\sigma}c^{\ph}_{i\sigma}\right)-\mu\sum_{i,\sigma}n_{i\sigma}\right]}\right)\right],
	\end{split}
\end{equation}
where $\text{Tr}$ acts exclusively over fermionic degrees of freedom. Note that the two different spins share the same auxiliary field configuration. The trace is over an action with only quadratic fermionic operators and can be explicitly carried out for a fixed configuration $C$ of the auxiliary fields:
\begin{equation}
	\label{eq:determinant}
	\begin{split}
	\text{Tr}&\left[\prod_\tau\left( e^{\sqrt{-\Delta\tau\frac{U}{2}}\sum_{i,\sigma}\eta_{i,\tau}\left(c^\dagger_{i\sigma}c^{\ph}_{i\sigma}-\frac{1}{2}\right)} e^{-\Delta\tau\left[\sum_{ij,\sigma}t_{ij,\sigma}\left(c^\dagger_{i\sigma}c^{\ph}_{j\sigma}+c^\dagger_{j\sigma}c^{\ph}_{i\sigma}\right)-\mu\sum_{i,\sigma}n_{i\sigma}\right]}\right)\right]\\
		&=\left[e^{-\sum_\tau\sum_i\sqrt{-\Delta\tau\frac{U}{2}}\eta_{i,\tau}/2}\right]^2\,\left[\text{det}\,\left(\mathbb{I}+\prod_\tau e^{\sqrt{-\Delta\tau\frac{U}{2}}\hat{V}} e^{-\Delta\tau\hat{T}}\right)\right]^2,
	\end{split}
\end{equation}
where the square in the last line comes from the spin degree of freedom and the single-particle matrices are defined as $[\hat{V}]_{ij}=\delta_{i,j}\eta_{i,\tau}$ and $[\hat{T}]_{ij}=t_{ij}-\mu\delta_{ij}$.

Whenever $U<0$, i.e., an attractive Hubbard model, the determinant is strictly real and its square is a positive real number. It is then possible to sample the auxiliary-field configurations with Monte Carlo techniques and treat $e^{-S(C)}$ as a probability weight.
For example, the expectation value of an observable $O$ would be given by
\begin{equation}
	\langle O \rangle=\frac{\text{Tr}\left[Oe^{-\beta H}\right]}{\text{Tr}\left[e^{-\beta H}\right]}=\sum_CP(C)\langle\langle O \rangle\rangle_C,
\end{equation}
where $P(C)=e^{-S(C)}/\sum_Ce^{-S(C)}$ and $\langle\langle \cdot \rangle\rangle_C$ indicates expectation value over the auxiliary-field configuration $C$.

New configurations are accepted with a Metropolis-Hastings acceptance probability. The ALF package considers sequential single spin flips. A discrete HS field at site $i$ and imaginary time $\tau$, $l_{i\tau}$ is randomly chosen and one of the three other possible values among $\pm 1, \pm 2$ is proposed with  probability $1/3$ \cite{Bercx:2017}. The choice of local updates has important consequences on the autocorrelation times. To estimate the autocorrelation time of different observables, we perform a rebinning analysis \cite{Ambegaokar:2010}. Given a series of $M$ measurements $O_i^{(0)}=O_i$, we iteratively join them in bins of larger sizes $O^{(n)}_i=\left(O^{(n-1)}_{2i-1}+O^{(n-1)}_{2i}\right)/2$ with $i\in[1,M/2n]$. We then compute the mean and variance of this measurement series assuming perfectly uncorrelated samples:
\begin{equation}
	\langle O\rangle=\sum_i\frac{O^{(n)}_i}{M/2^n},
\end{equation}
and 
\begin{equation}
	\text{Var}_{O^{(n)}}=\sqrt{\frac{1}{M/2^n(M/2^n-1)}\sum_i\left(O^{(n)}_i-\langle O^{(n)}\rangle \right)^2}.
\end{equation}
$\text{Var}_{O^{(n)}}$ converges to the correct unbiased error estimate for $n\to \infty$. Therefore, we gradually increase the bin size until the error converges as a function of the bin's size. 

Finally, let us touch upon how to obtain the expectation values of interest from these Monte Carlo simulations. One appealing feature of the auxiliary-field quantum Monte Carlo method is that the interacting fermionic problem gets mapped to a non-interacting one via HS transformation. Therefore, at each fixed configuration $C$ of the auxiliary fields, we deal with free fermions. Hence, Wick's theorem applies and allows to readily obtain any expectation value starting from the two-body expectation value of the bare Green's function: $\langle c^\dagger_{j\sigma} c^{\ph}_{i\sigma}\rangle$. Note that here $\langle \cdot \rangle$ indicates averages on a fixed configuration $C$ of the auxiliary fields. For example,
\begin{equation}
	\langle c^\dagger_\alpha c^\dagger_\beta c^{\ph}_\delta c^{\ph}_\gamma \rangle = \langle  c^\dagger_\alpha  c^{\ph}_\gamma\rangle \langle c^\dagger_\beta c^{\ph}_\delta\rangle -  \langle  c^\dagger_\alpha  c^{\ph}_\delta\rangle \langle c^\dagger_\beta c^{\ph}_\gamma\rangle ,
\end{equation}
where the minus sign comes for fermionic anti-commutation relations. 
Thanks to the SU(2) symmetry, we also have: $\langle c^\dagger_{j\uparrow} c^{\ph}_{i\uparrow}\rangle$=$\langle c^\dagger_{j\downarrow} c^{\ph}_{i\downarrow}\rangle$. The expectation value of equal-time correlators are then obtained as follows:
\begingroup
\allowdisplaybreaks
\begin{align}
	\text{Kinetic energy}:\quad E_K&=2\sum_{ij}t_{ij}\langle c^\dagger_{i}c^{\ph}_{j} \rangle, \\
	\begin{split}
	\text{Interacting energy}:\quad E_I&=U\sum_i\langle c^\dagger_{i\uparrow}c^{\ph}_{i\uparrow} c^\dagger_{i\downarrow}c^{\ph}_{i\downarrow} \rangle \\ & = U\sum_i\langle c^\dagger_{i\uparrow}c^{\ph}_{i\uparrow}\rangle\langle c^\dagger_{i\downarrow}c^{\ph}_{i\downarrow} \rangle = U\sum_i\langle c^\dagger_{i}c^{\ph}_{i} \rangle^2,
	\end{split}\\
	\text{Particle number}:\quad N&=2\sum_i\langle c^\dagger_{i}c^{\ph}_{i} \rangle, \\
	\text{Diamagnetic current}:\quad \langle T_{\mu\nu} \rangle &= 2\sum_{ij} \frac{\partial^2 t_{ij}(\mathbf{r})}{\partial r_\mu\partial r_\nu}\langle c^\dagger_{i}c^{\ph}_{j} \rangle,\\\begin{split}
	z-\text{Spin correlator}:\quad \langle S^z_iS^z_j \rangle&=\sum_{ij} \left\langle \left(c^\dagger_{i\uparrow}c^{\ph}_{i\uparrow}-c^\dagger_{i\downarrow}c^{\ph}_{i\downarrow}\right)\left(c^\dagger_{j\uparrow}c^{\ph}_{j\uparrow}-c^\dagger_{j\downarrow}c^{\ph}_{j\downarrow}\right) \right\rangle\\
	& = \sum_{ij} 2\langle c^\dagger_{i}c^{\ph}_{i} \rangle\langle c^\dagger_{j}c^{\ph}_{j}\rangle + 2\langle c^\dagger_{i}c^{\ph}_{j}\rangle\langle c^{\ph}_{i}c^\dagger_{j} \rangle  - 2\langle c^\dagger_{i\uparrow}c^{\ph}_{i} \rangle\langle c^\dagger_{j}c^{\ph}_{j} \rangle\\
	& = 2\sum_{ij} \langle c^\dagger_{i}c^{\ph}_{j}\rangle\langle c^{\ph}_{i}c^\dagger_{j} \rangle,  \end{split}\\ \begin{split}
	\text{Density correlator}:\quad \langle N_iN_j \rangle&= \sum_{ij} \left\langle \left(c^\dagger_{i\uparrow}c^{\ph}_{i\uparrow}+c^\dagger_{i\downarrow}c^{\ph}_{i\downarrow}\right)\left(c^\dagger_{j\uparrow}c^{\ph}_{j\uparrow}+c^\dagger_{j\downarrow}c^{\ph}_{j\downarrow}\right)\right\rangle \\
	&  = \sum_{ij} 4\langle c^\dagger_{i}c^{\ph}_{i} \rangle\langle c^\dagger_{j}c^{\ph}_{j}\rangle + 2 \langle c^\dagger_{i}c^{\ph}_{j} \rangle\langle c^{\ph}_{i}c^\dagger_{j}\rangle,
	\end{split}\\ \begin{split}
	\text{Pair correlator}:\quad  \langle \Delta_i\Delta_j \rangle&=
	\sum_{ij}\left\langle \left(c^{\ph}_{i\uparrow}c^{\ph}_{i\downarrow}+c^\dagger_{i\downarrow}c^\dagger_{i\uparrow}\right)\left(c^{\ph}_{j\uparrow}c^{\ph}_{j\downarrow}+c^\dagger_{j\downarrow}c^\dagger_{j\uparrow}\right)\right\rangle\\
	& =\sum_{ij} \langle c^{\ph}_{i\uparrow}c^{\ph}_{i\downarrow}c^\dagger_{j\downarrow}c^\dagger_{j\uparrow} \rangle + \langle c^\dagger_{i\downarrow}c^\dagger_{i\uparrow}c^{\ph}_{j\uparrow}c^{\ph}_{j\downarrow} \rangle \\
	& = \sum_{ij} \langle c^{\ph}_i c^\dagger_j \rangle^2 + \langle c^\dagger_i c^{\ph}_j \rangle^2.
	\end{split}\\
\end{align}
\endgroup
The calculation of the current-current correlator
\begin{equation}
	\Lambda_{\mu\nu}^{\alpha\beta}(i,j)=\langle J_\mu^\alpha(i) J_\nu^\beta(j) \rangle,
\end{equation}
with $J_\mu^\alpha(i)=\sum_\sigma f_\mu^\alpha c^\dagger_{i_1\sigma}c^{\ph}_{i_2\sigma} + (f_\mu^\alpha)^* c^\dagger_{i_2\sigma}c^{\ph}_{i_1\sigma}$ is more complicate. Here $i$ and $j$ are unit cell indices, while $\alpha$ ($\beta$) is the bond between orbitals $i_1$ ($j_1$) and $i_2$ ($j_2$). Finally, $f_\mu^{\alpha}=\partial t_\alpha(\mathbf{r})/\partial r_\mu$.

\begin{equation}
	\begin{split}
		\Lambda_{\mu\nu}^{\alpha\beta}(i,j)=&\sum_{\sigma\sigma'}f_\mu^\alpha f_\nu^\beta\langle c^\dagger_{i_1\sigma}c^{\ph}_{i_2\sigma} c^\dagger_{j_1\sigma'}c^{\ph}_{j_2\sigma'} \rangle + f_\mu^\alpha (f_\nu^\beta)^*\langle c^\dagger_{i_1\sigma}c^{\ph}_{i_2\sigma} c^\dagger_{j_2\sigma'}c^{\ph}_{j_1\sigma'} \rangle +\\&\quad (f_\mu^\alpha)^* f_\nu^\beta\langle c^\dagger_{i_2\sigma}c^{\ph}_{i_1\sigma} c^\dagger_{j_1\sigma'}c^{\ph}_{j_2\sigma'}  \rangle + (f_\mu^\alpha)^* (f_\nu^\beta)^*\langle c^\dagger_{i_2\sigma}c^{\ph}_{i_1\sigma} c^\dagger_{j_2\sigma'}c^{\ph}_{j_1\sigma'} \rangle \\=&
		\sum_{\sigma\sigma'} f_\mu^\alpha f_\nu^\beta \left(\langle c^\dagger_{i_1\sigma}c^{\ph}_{i_2\sigma} \rangle \langle c^\dagger_{j_1\sigma'}c^{\ph}_{j_2\sigma'} \rangle + \delta_{\sigma\sigma'} \langle c^\dagger_{i_1\sigma}c^{\ph}_{j_2\sigma'}  \rangle \langle c^{\ph}_{i_2\sigma}c^\dagger_{j_1\sigma'} \rangle\right) +\\
		&\quad f_\mu^\alpha (f_\nu^\beta)^* \left(\langle c^\dagger_{i_1\sigma}c^{\ph}_{i_2\sigma} \rangle \langle c^\dagger_{j_2\sigma'}c^{\ph}_{j_1\sigma'} \rangle + \delta_{\sigma\sigma'} \langle c^\dagger_{i_1\sigma}c^{\ph}_{j_1\sigma'}  \rangle \langle c^{\ph}_{i_2\sigma}c^\dagger_{j_2\sigma'} \rangle\right) +\\
		& \quad (f_\mu^\alpha)^* f_\nu^\beta \left(\langle c^\dagger_{i_2\sigma}c^{\ph}_{i_1\sigma} \rangle \langle c^\dagger_{j_1\sigma'}c^{\ph}_{j_2\sigma'} \rangle + \delta_{\sigma\sigma'} \langle c^\dagger_{i_2\sigma}c^{\ph}_{j_2\sigma'}  \rangle \langle c^{\ph}_{i_1\sigma}c^\dagger_{j_1\sigma'} \rangle\right)+\\
		& \quad (f_\mu^\alpha)^* (f_\nu^\beta)^* \left(\langle c^\dagger_{i_2\sigma}c^{\ph}_{i_1\sigma} \rangle \langle c^\dagger_{j_2\sigma'}c^{\ph}_{j_1\sigma'} \rangle + \delta_{\sigma\sigma'} \langle c^\dagger_{i_2\sigma}c^{\ph}_{j_1\sigma'}  \rangle \langle c^{\ph}_{i_1\sigma}c^\dagger_{j_2\sigma'} \rangle\right) \\=	
		& f_\mu^\alpha f_\nu^\beta \left(4\langle c^\dagger_{i_1}c^{\ph}_{i_2} \rangle \langle c^\dagger_{j_1}c^{\ph}_{j_2} \rangle + 2 \langle c^\dagger_{i_1}c^{\ph}_{j_2}  \rangle \langle c^{\ph}_{i_2}c^\dagger_{j_1} \rangle\right) +\\
				&\quad f_\mu^\alpha (f_\nu^\beta)^* \left(4\langle c^\dagger_{i_1}c^{\ph}_{i_2} \rangle \langle c^\dagger_{j_2}c^{\ph}_{j_1} \rangle + 2\langle c^\dagger_{i_1}c^{\ph}_{j_1}  \rangle \langle c^{\ph}_{i_2}c^\dagger_{j_2} \rangle\right) +\\
				& \quad (f_\mu^\alpha)^* f_\nu^\beta \left(4\langle c^\dagger_{i_2}c^{\ph}_{i_1} \rangle \langle c^\dagger_{j_1}c^{\ph}_{j_2} \rangle + 2 \langle c^\dagger_{i_2}c^{\ph}_{j_2}  \rangle \langle c^{\ph}_{i_1}c^\dagger_{j_1} \rangle\right)+\\
				& \quad (f_\mu^\alpha)^* (f_\nu^\beta)^* \left(4\langle c^\dagger_{i_2}c^{\ph}_{i_1} \rangle \langle c^\dagger_{j_2}c^{\ph}_{j_1} \rangle + 2 \langle c^\dagger_{i_2}c^{\ph}_{j_1}  \rangle \langle c^{\ph}_{i_1}c^\dagger_{j_2} \rangle\right). 
	\end{split}
\end{equation}

Time-displaced correlators can be readily obtained via the following substitutions: $c_i\to c_i(\tau)$ and $c_j\to c_j(0)$, while for the current-current correlator: $c_{i1}\to c_{i1}(\tau)$, $c_{j1}\to c_{j1}(0)$, $c_{i2}\to c_{i2}(\tau)$ and $c_{j2}\to c_{j2}(0)$.

Momentum space correlators as those of Eq.~\eqref{eq:corrcorr} are obtained via Fourier transform:
\begin{equation}
	C(\mathbf{q})=\frac{1}{N}\sum_{ij}e^{-i\mathbf{q}\cdot(\mathbf{r}_i-\mathbf{r}_j)}C(\mathbf{r}_i-\mathbf{r}_i).
\end{equation}
Extra care needs to be taken for the current-current correlator, since the current operator is a bond operator. We first Fourier transform as
\begin{equation}
	\Lambda_{\mu\nu}^{\alpha\beta}(\mathbf{q})=\sum_{ij}e^{-i\mathbf{q}\cdot(\mathbf{r}_i-\mathbf{r}_j)}\int_0^\beta\text{d}\tau\langle J_\mu^\alpha(i,\tau) J_\nu^\beta(j,0) \rangle.
\end{equation}
Only then we sum these components keeping track of the bonds's location $\mathbf{d}_{\alpha}=\mathbf{r}_{i2}-\mathbf{r}_{i1}$:
\begin{equation}
	\Lambda_{\mu\nu}(\mathbf{q})=\sum_{\alpha\beta}e^{-i\mathbf{q}\cdot(\mathbf{d}_\alpha-\mathbf{d}_\beta)}\Lambda_{\mu\nu}^{\alpha\beta}(\mathbf{q}).
\end{equation}

In this study we use a Trotter discretization $\Delta\tau=0.1$. Results are averages over 50--100 independent runs, each made of 150 sweeps. Here, a single sweep is such that each auxiliary field is visited twice in a sequential propagation from $\tau = 0$ to $\tau=\beta/\Delta \tau$ and backwards \cite{Bercx:2017}.

Note that also the nearest-neighbor interaction studied in Sec.~\ref{sec:DSusc} can be written as a square of quadratic terms in fermionic operators and lends itself to simulations via AFQMC. We consider the interaction:
\begin{equation}
	\begin{split}
	H_{V}&= V \sum_{\langle i,j \rangle}\left[\sum_\sigma\left(c_{i,\sigma}^\dagger c_{i,\sigma}-1/2\right)\right]\left[\sum_{\sigma'}\left(c_{j,\sigma'}^\dagger c_{j,\sigma'}-1/2\right)\right]= V \sum_{\langle i,j \rangle}\left(n_i-1\right)\left(n_j-1\right)\\ & =\frac{V}{2}\sum_{\langle i,j \rangle} \left[\left(n_i+n_j-2\right)^2-\left(n_j-1\right)^2-\left(n_i-1\right)^2\right],
	\end{split}
\end{equation} 
with $V<0$.
Note that the addition of the nearest-neighbor interaction leads to a renormalization of the on-site density-density interaction: $U'=U-zV$, where $z=4$ is the number of nearest-neighbors for each lattice site.

\section{Single-particle density of states}
Fig. 3(a) of the main text shows the temperature dependence of the single-particle density of states $N(\epsilon_F)$ defined as:
\begin{equation}
	\label{eq:dos}
	N(\epsilon_F)=\frac{\beta}{\pi L^2}\sum_{\alpha}\int_\text{BZ}d{\bf k}\,\langle c^{\phantom{\dagger}}_{\alpha\mathbf{k}}(\beta/2)c^\dagger_{\alpha\mathbf{k}}(0)\rangle.
	\end{equation}
The single-particle density of states at arbitrary energy $\omega$ (measured from $\mu$) is given by the spectral function $A(\mathbf{k},\omega)$: $N(\omega)=1/L^2\int_{BZ} d^2 k\, A(\mathbf{k},\omega)$. In our simulations, we rather have access to the imaginary time Green's function $G(\mathbf{k},\tau)=\sum_\alpha\langle c_{\alpha\mathbf{k}}(\tau)c^\dagger_{\alpha\mathbf{k}}(0)\rangle$, where $\alpha$ is the sublattice index.  $G(\mathbf{k},\tau)$ is related to the spectral function by
	\begin{equation}
		\label{eq:proxyA}
	G(\mathbf{k},\tau)=\int_{-\infty}^{+\infty} d\omega\frac{e^{-\omega\tau}}{1+e^{-\beta\omega}}A(\mathbf{k},\omega),
\end{equation}
for $0 < \tau < \beta$.
Note that $G(\mathbf{k},\beta/2)$ is the integrated spectral weight around a window of width $\sim T$ and is directly related to the single-electron spectral function: $\pi A(\mathbf{k},\epsilon_F)=\lim_{T\to 0}\beta G(\mathbf{k},\beta/2)$. This fact allows us to avoid inverting the Laplace transform of Eq.~\eqref{eq:proxyA}.
Namely, to probe the opening of a gap in the single-particle spectrum, we can simply use Eq. \eqref{eq:dos} \cite{Trivedi:1995}.

\begin{figure}[t]
\includegraphics[]{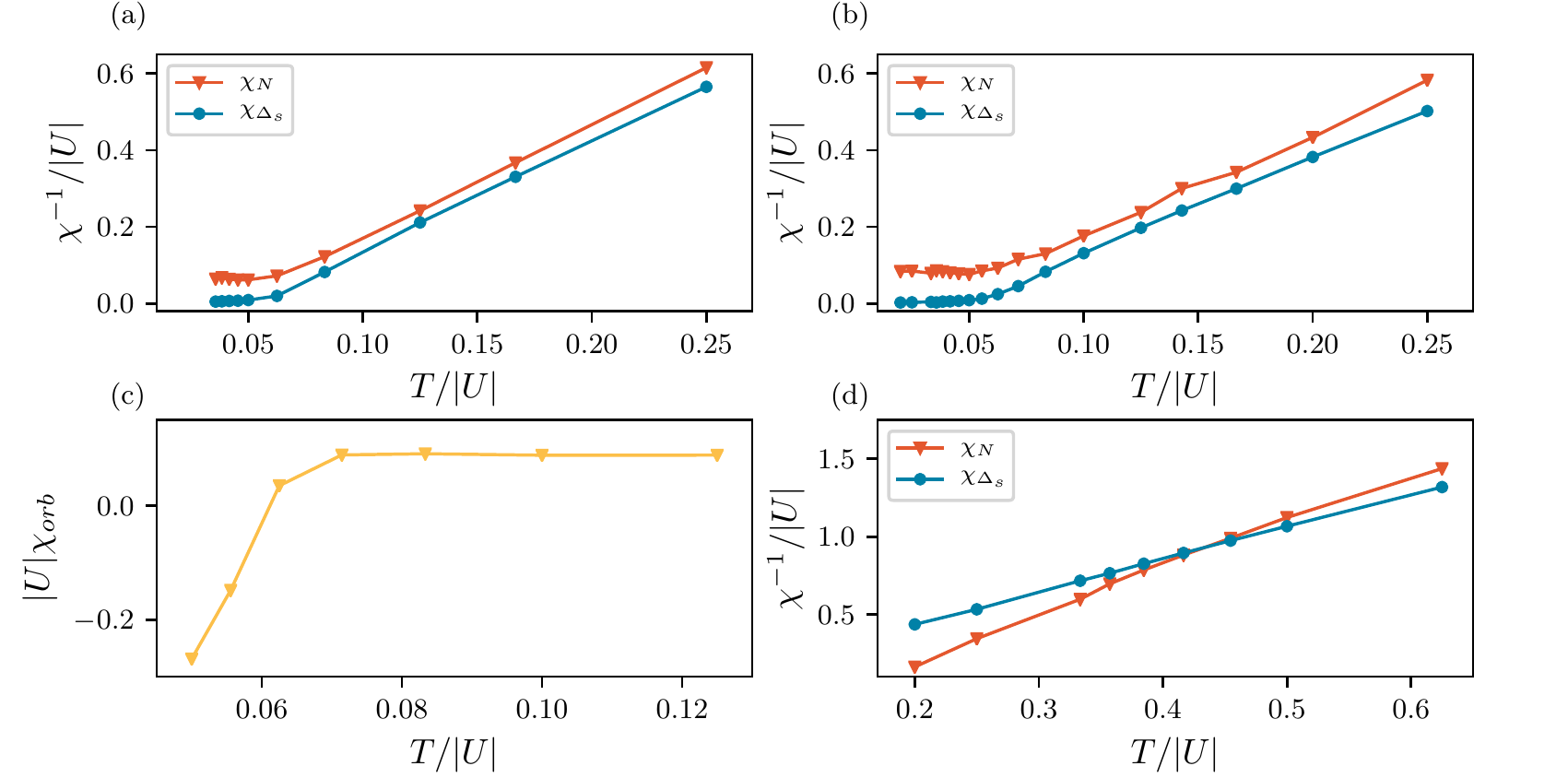}
\caption{\label{fig:fig2supp} Density $\chi_N$ and pair susceptibility $\chi_{\Delta_s}$ for the kagome-3 model with $L=6$ and (a) $\lvert U\rvert=1$, (b) $\lvert U\rvert=2$. $\chi_N$ peaks at $T\approx T_N$ reflecting a tendency towards phase separation that is suppressed by the onset of superconductivity. (c) Orbital susceptibility $\chi_{orb}$ for the kagome-3 model with $L=8$ and $\lvert U\rvert=2$. It turns negative at $T/\lvert U \rvert \approx 0.06$ signaling the onset of diamagnetism. (d) Density $\chi_N$ and pair susceptibility $\chi_{\Delta_s}$ for the kagome-3 model with $L=6$, $\lvert U\rvert=2$ and additional nearest-neighbor interaction $\lvert V\rvert=0.2$. $\chi_N$ gets larger than $\chi_{\Delta_s}$, and phase separation prevails.}
\end{figure}
\section{Density and pair susceptibilities}
\label{sec:DSusc}
We study the density and pair susceptibilities
\begin{equation}
	 \label{eq:suscusc}
 	\chi_{\mathcal{O}}=\frac{1}{L^2}\int_0^\beta d\tau \langle \mathcal{O}(\tau)\mathcal{O}(0) \rangle ,
 \end{equation}
with $\mathcal{O}$ being the charge, $N=\sum_i(n_i-\nu)$, and $s$-wave pairing, $\Delta_s=\sum_i c^{\phantom{\dagger}}_{i\uparrow}c^{\phantom{\dagger}}_{i\downarrow}+c^\dagger_{i\downarrow}c^\dagger_{i\uparrow}$, operators.  

Both $\chi_N$ and $\chi_{\Delta_s}$ are greatly enhanced at low temperatures and follow a similar trend up to $T\approx T_N$, where $\chi_N$ reaches a maximum and seems to saturate, cf. Figs. \ref{fig:fig2supp}(a)--(b). $\chi_{\Delta_s}$, instead, continues to grow. The great enhancement of $\chi_N$ up to $T_N$ is linked to the tendency towards phase separation due to an emergent SU(2) symmetry in the limit $\lvert U \rvert/\delta \to 0$ \cite{Tovmasyan:2016}. The role of this emergent SU(2) symmetry is carefully discussed in Ref.~[\onlinecite{Tovmasyan:2016}] and the Supplemental Material of Ref.~[\onlinecite{Hofmann:2019}]. Here we briefly report their main conclusions. 

Consider a single exactly flat band satisfying the uniform pairing condition of Sec. \ref{ssec:uniform}. Upon projection on the active flat band, the purely interacting Hubbard Hamiltonian possesses an emergent SU(2) symmetry. The generators of this symmetry are $\tau^z=\frac{1}{2}\sum_{\mathbf{k},\sigma}c^\dagger_{\mathbf{k},\sigma}c^{\phantom{\dagger}}_{\mathbf{k},\sigma}$, $\tau^+=\sum_{\mathbf{k}}c^\dagger_{\mathbf{k},\uparrow}c^\dagger_{-\mathbf{k},\downarrow}$, and $\tau^-=(\tau^+)^\dagger$. These generators show how $s$-wave pair and density susceptibilities are identical in the presence of an exact SU(2) symmetry. Moreover, the Hohenberg-Mermin-Wagner theorem prevents a finite $T_c$ in two-dimensional systems if the SU(2) symmetry is exact.

Multiple effects can break the emergent SU(2) symmetry. In our studies, it is broken by a finite gap $\lvert U \rvert / \delta \not \to 0$ and finite temperatures. Our numerical results confirm that a finite gap helps to stabilize the superconducting state, in accordance with the findings of Refs. [\onlinecite{Tovmasyan:2016}] and [\onlinecite{Hofmann:2019}].

One might wonder whether different interactions favor a phase separated ground state. We consider an additional term in the Hamiltonian of Eq.~\eqref{eq:wholeH}: 
\begin{equation}
	H_{V}=-\lvert V \rvert \sum_{\langle i,j \rangle}\left[\sum_\sigma\left(c_{i,\sigma}^\dagger c_{i,\sigma}-1/2\right)\right]\left[\sum_{\sigma'}\left(c_{j,\sigma'}^\dagger c_{j,\sigma'}-1/2\right)\right],
\end{equation}
where the sum runs over nearest-neighbors bonds. Even for small $V=0.1 U$, the density susceptibility $\chi_N$ is enhanced and at low temperatures gets bigger than the pair susceptibility $\chi_{\Delta_s}$, cf. Fig.~\ref{fig:fig2supp}(d). This finding shows that nearest-neighbor interactions favor a phase separated ground state in this flat-band model. 

\begin{figure}[t]
\includegraphics[]{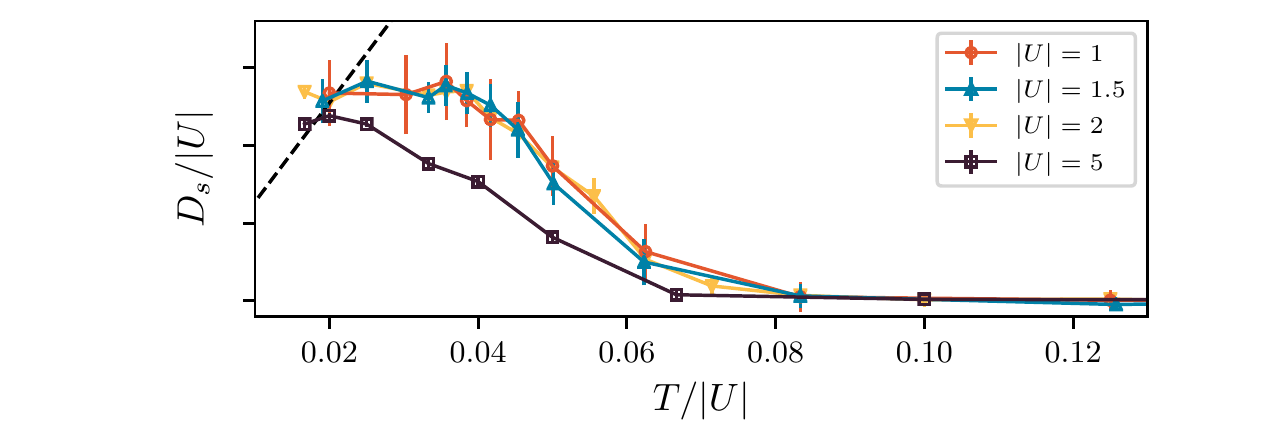}
\caption{\label{fig:fig4supp} 
Superfluid weight $D_s(T)$ for the attractive Hubbard model with interaction strength $\lvert U \rvert$. The crossing of $D_s$ with the dashed line $2T/\pi$ indicates the BKT transition, where the superconducting transition occurs. Different interaction strengths $\lvert U \rvert = 1,\, 1.5,\, 2,\, 5$ in a 6 × 6 system.}
\end{figure}

\section{Orbital susceptibility}
The orbital magnetic susceptibility allows distinguishing between a paramagnetic and a diamagnetic behavior. It is defined as:
\begin{equation}
	\chi_{orb}=\lim_{k\to 0}\frac{1}{k^2}\left[\Lambda_{xx}(0,k)-\Lambda_{xx}(k,0)\right],
	\end{equation}
where $\Lambda_{xx}$ is the current-current correlator of Eq. (8) of the main text. Fig. \ref{fig:fig2supp}(c) shows how $\chi_{\text{orb}}$ turns negative for $T/\lvert U \rvert \approx 0.06$, before the BKT transition. The onset of pairing fluctuations might indeed explain the diamagnetic behavior observed above $T_c$. 

\section{Beyond the isolated flat band limit}
In the main text, we exclusively consider interactions $\lvert U \rvert < \delta$. In this range, the linear scaling of $T_c$ with $\lvert U \rvert$ continued for values that violate the isolated flat band condition: $\lvert U \rvert \ll \delta$. 

We expect this linear relation to stop once $\lvert U \rvert $ exceeds the energy gap to the dispersive upper band. This intuition is substantiated by Monte Carlo simulations for $\lvert U \rvert=5$. As shown in Fig.~\ref{fig:fig4supp}, the superfluid weight scaled with respect to $\lvert U \rvert$ does not fall on top of the traces for $\lvert U \rvert < \delta$ and $T_c<0.02\lvert U\rvert$. 

\section{Competing orders}
To monitor the possible presence of competing orders we study the equal-time correlator for the operator $\mathcal{O}$:
\begin{equation} \label{eq:corrcorr} C_\mathcal{O}(k)=\frac{1}{L^2}\sum_{ij}e^{-i(\mathbf{r}_i-\mathbf{r}_j)\cdot \mathbf{k}}\langle \mathcal{O}^\dagger(\mathbf{r}_i)\mathcal{O}(\mathbf{r}_j)  \rangle.
	\end{equation}
In particular, as operator $\mathcal{O}$, we consider the spin $S_z=\sum_i \left( n_{i\uparrow}-n_{i\downarrow}\right)$, charge $N=\sum_i \left( n_{i\uparrow}+n_{i\downarrow}\right)$ and pair $\Delta_s=\sum_i \left( c^{\phantom{\dagger}}_{i\uparrow}c^{\phantom{\dagger}}_{i\downarrow}+c^\dagger_{i\downarrow}c^\dagger_{i\uparrow}\right)$ operators.

\begin{figure}[t]
\includegraphics[]{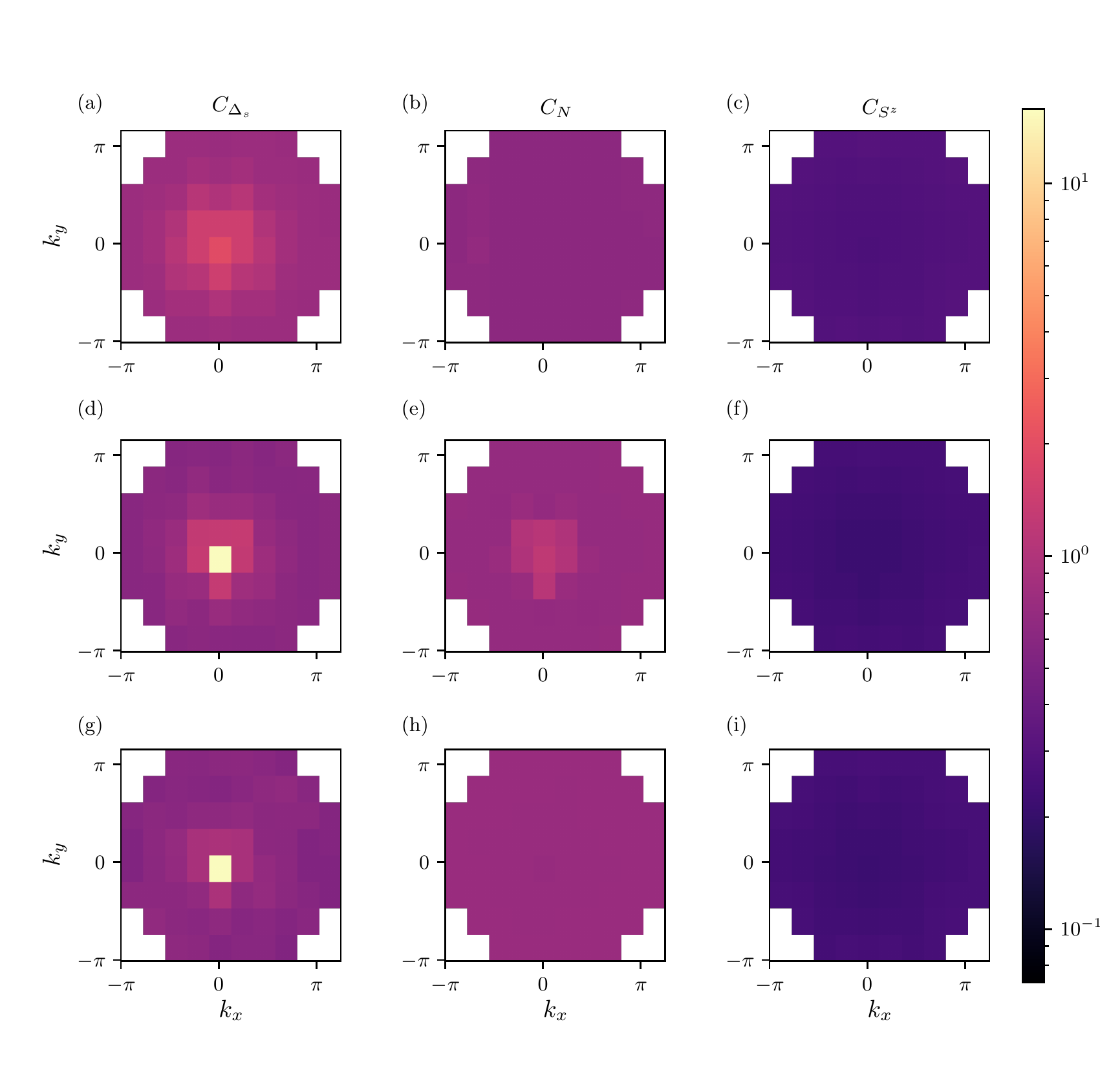}
\caption{\label{fig:fig3supp} In the first column the pair correlator $C_{\Delta_s}$, in the second the density correlator $C_N$ and in the last the spin correlator $C_{S^z}$. The first row is at temperature $T\approx 2T_c$, the second at $T\approx T_c$ and the third at $T\approx T_c/2$. All plots refer to a kagome-3 model with $L=8$ and $\lvert U \rvert=2$. At $T\approx T_c$ a clear onset of phase coherence is signaled by a peak at $\mathbf{k}=0$ of $C_{\Delta_s}$. The small enhancement at $\mathbf{k}=0$ of $C_N$ confirms a tendency towards phase separation, suppressed by the onset of superconductivity. Finally $C_{S^z}$ appears featureless across the whole temperature range.}
\end{figure}

$C_{S^z}$ looks featureless across all temperature scales. $C_N$ shows a peak at $\mathbf{k}=0$ around $T_c$ supporting the claim of a tendency towards phase separation due to an approximate emergent SU(2) symmetry. $C_{\Delta_s}$ develops a strong peak at $\mathbf{k}=0$ for $T\leq T_c$ signaling the onset of phase coherence in the pair formation and the transition to a superconducting state. The peak in $C_N$ disappears for $T< T_c$, proving the prevalence of the superconducting order over phase separation. All these correlators are shown in Fig. \ref{fig:fig3supp} for temperatures $T\approx 2 T_c$, $T\approx T_c$, and $T\approx T_c/2$.
\newpage

%

	\end{bibunit}

	\end{document}